\documentclass[12pt]{article}
\usepackage{amsmath,amssymb}
\usepackage{bm}

\catcode`\@=11

\ifx\documentclass\undefined
 \def\@sect#1#2#3#4#5#6[#7]#8{\ifnum #2>\c@secnumdepth
     \let\@svsec\@empty\else
     \refstepcounter{#1}\edef\@svsec{\csname prefix#1\endcsname
        \csname the#1\endcsname\hskip 1em}\fi
     \@tempskipa #5\relax
      \ifdim \@tempskipa>\z@
        \begingroup #6\relax
          \@hangfrom{\hskip #3\relax\@svsec}{\interlinepenalty \@M #8\par}%
        \endgroup
       \csname #1mark\endcsname{#7}\addcontentsline
         {toc}{#1}{\ifnum #2>\c@secnumdepth \else
                      \protect\numberline{\csname the#1\endcsname}\fi
                    #7}\else
        \def\@svsechd{#6\hskip #3\relax  
                   \@svsec #8\csname #1mark\endcsname
                      {#7}\addcontentsline
                           {toc}{#1}{\ifnum #2>\c@secnumdepth \else
                             \protect\numberline{\csname the#1\endcsname}\fi
                       #7}}\fi
     \@xsect{#5}}
\else
    \def\@seccntformat#1{\csname prefix#1\endcsname
        \csname the#1\endcsname\quad}
\fi

\@addtoreset{equation}{section}   
\def\theequation{\arabic{section}.\arabic{equation}}

\def\thebibliography#1{\section*{References\@mkboth
 {REFERENCES}{REFERENCES}}\list
 {\leftbibmark\arabic{enumi}\rightbibmark}{
 \settowidth\labelwidth{\leftbibmark #1\rightbibmark}\leftmargin\labelwidth
 \advance\leftmargin\labelsep
 \usecounter{enumi}}
 \def\newblock{\hskip .11em plus .33em minus -.07em}
 \sloppy\clubpenalty4000\widowpenalty4000
 \sfcode`\.=1000\relax}

\def\@citex[#1]#2{\if@filesw\immediate\write\@auxout{\string\citation{#2}}\fi
  \def\@citea{}\@cite{\@for\@citeb:=#2\do
    {\@citea\def\@citea{,\penalty\@m\ }\@ifundefined
       {b@\@citeb}{{\bf ?}\@warning
       {Citation `\@citeb' on page \thepage \space undefined}}%
\hbox{\csname b@\@citeb\endcsname\citemarkdelim}}}{#1}}

\def\@cite#1#2{\leftcitemark{#1 \if@tempswa , #2\fi}\rightcitemark}
\def\leftcitemark{[}
\def\rightcitemark{]}
\def\citemarkdelim{}
\def\leftbibmark{[}
\def\rightbibmark{]}

\catcode`\@=12

\usepackage{amsmath}
\usepackage{amssymb}

\begin{document}

\begin{titlepage}

\vspace{2cm}

\begin{center}\large\bf
Long wavelength limit of evolution of nonlinear 
cosmological perturbations\\ 
\end{center}

\begin{center}
Takashi Hamazaki\footnote{email address: yj4t-hmzk@asahi-net.or.jp}
\end{center}

\begin{center}\it
Kamiyugi 3-3-4-606 Hachioji-city\\
Tokyo 192-0373 Japan\\
\end{center}

\begin{center}\bf Abstract\end{center}

In the general matter composition where the multiple scalar fields
and the multiple perfect fluids coexist, in the leading order of 
the gradient expansion, we construct all the solutions of the 
nonlinear evolutions of the locally homogeneous universe.
From the momentum constraint, we derive the constraints which the 
solution constants of the locally homogeneous universe must 
satisfy. 
We construct the gauge invariant perturbation variables 
in the arbitrarily higher order nonlinear cosmological perturbation 
theory around the spatially flat FRW universe.
We construct the nonlinear LWL formula representing the long 
wavelength limit of the evolution of the nonlinear
gauge invariant perturbation variables in terms of 
perturbations of the evolutions of the locally homogeneous 
universe.
By using the LWL formula, we investigate the evolution of nonlinear 
cosmological perturbations in the universe dominated by the multiple
slow rolling scalar fields with an arbitrary potential.
The $\tau$ function and the $N$ potential introduced in this paper make
it possible to write the evolution of the multiple slow rolling 
scalar fields with an arbitrary interaction potential and the
arbitrarily higher order nonlinear Bardeen parameter at the end of 
the slow rolling phase analytically.
It is shown that the nonlinear parameters such as $f_{NL}$, $g_{NL}$
are suppressed by the slow rolling expansion parameters.

\vspace{1cm}

  
PACS number(s):98.80.Cq

\end{titlepage}

\section{Introduction and summary}

In the inflationary universe scenario, the quantum fluctuations
of the scalar fields called inflatons are thought to be the origin
of the cosmological large scale structures such as galaxies and 
clusters of galaxies.
In the slow rolling phase, the quantum fluctuations are stretched 
into the superhorizon scales and stay outside the horizon until
they return into the horizon in the radiation dominant universe.
Therefore the method for investigating the evolutions of the 
long wavelength cosmological perturbations became necessary and
the LWL method was developed \cite{Taruya1998} \cite{Kodama1998} 
\cite{Sasaki1998}.
In the LWL method, we use the LWL formulae representing the long
wavelength limit of the evolutions of the cosmological perturbations 
in terms of the quantities of the corresponding exactly homogeneous 
universe.
As for the adiabatic modes the exact LWL formula had already been
constructed and it was used in order to investigate the slow 
rolling phase \cite{Polarski1992} and the oscillatory phase
\cite{Kodama1996} \cite{Hamazaki1996}.
Nambu and Taruya suggested in the multiple scalar fields system also, 
the LWL formula exists \cite{Taruya1998}. 
Soon in this case the exact LWL formula was constructed \cite{Kodama1998} 
\cite{Sasaki1998} and it was used to investigate the multiple
oscillatory scalar fields \cite{Hamazaki2002} \cite{Hamazaki2004}.
In addition, in the paper \cite{Kodama1998}, we presented the flexible
way for constructing the LWL formulae in the general matter composition.
It will be called the Kodama and Hamazaki (KH) construction in this 
paper.
In the KH construction, the perturbation variables related with the 
exactly homogeneous universe, such as the scalar field perturbation
and the energy density perturbation, are expressed in terms of
the exactly homogeneous perturbations, that is the derivative of 
the exactly homogeneous quantity with respect to the solution 
constant.
By solving the spatial components of the Einstein equations,
the perturbation variables not related with the exactly 
homogeneous quantities such as vector quantities,
for example velocity perturbation variables, are expressed in the 
form of the integral of the perturbation variables related with
the exactly homogeneous universe which have already been determined.
We pointed out that the perturbation solution constants contained 
in the expressions of the perturbation variables determined in the 
processes explained above must satisfy the constraint coming from 
the momentum constraint.
The KH construction can be applied to the system containing 
the perfect fluid components having vector degrees 
of freedom such as the velocity perturbation variables.
Therefore by using the KH construction, 
in the most general matter composition
where multiple scalar fields and multiple perfect fluids coexist,
the LWL formula was constructed and used to investigate the multiple
component reheating and the multiple component curvaton decay
\cite{Hamazaki2008}.
Afterward in case of the nonlinear perturbation also, the existence
of the LWL formula was suggested \cite{Lyth2005} \cite{Rigopoulos2003}.
In this paper we give the definition of an arbitrarily higher order
nonlinear gauge invariant perturbation variables and 
the exact nonlinear LWL formula representing
them in terms of the derivatives of the quantities of the locally
homogeneous universe with respect to the solution constants.

In order to calculate the present density perturbations and the
cosmic microwave background anisotropies from the initial seed
perturbations, we need to calculate the evolution of the scalar 
fields perturbations on superhorizon scales during the slow rolling 
phase.
In the papers \cite{Gordon2001} \cite{Byrnes2006}, by decomposing
the multiple scalar fields into the adiabatic field and the entropy
fields instant by instant, the evolutionary behaviors of cosmological
perturbations during the slow rolling phase were discussed.
But this study is the local investigation, that is, it is based on
the Taylor expansion of the evolution equations around the first 
horizon crossing and the evolutions of the scalar fields in the 
long time interval were not solved.
Then in this paper we present the method which makes it possible 
to trace the evolutions of the multiple slow rolling scalar fields 
with an arbitrary interaction potential for the long time period 
as in the whole slow rolling phase.

This paper is organized as follows.
The section $2$ is devoted to the nonlinear LWL formula.  
In the subsection $2.1$, under the assumption that the universe
is the spatial flat FRW universe in the background level, 
in the leading order of the gradient expansion, 
we give the evolution equations of the locally homogeneous universe.
We point out that these locally homogeneous evolution equations 
are similar to the corresponding exactly homogeneous evolution 
equations and that deviations between the two are induced by the 
unimodular factor of the spatial metric $\tilde{\gamma}_{ij}$
which has contribution from the adiabatic decaying mode.
Following the philosophy of the KH construction \cite{Kodama1998},
we construct all the solutions of the evolution equations of 
the locally homogeneous universe.
We give the constraint which comes from the momentum constraint
and which must be satisfied by the solution constants contained
in the solution of the locally homogeneous universe.
In the subsection $2.2$, we give the definition of the gauge invariant
perturbation variables of the arbitrarily higher order nonlinear 
cosmological perturbation theory, and prove their gauge invariance.
We present the nonlinear LWL formula representing the long wavelength
limit of the evolutions of the gauge invariant perturbation variables 
in the nonlinear perturbation theory in terms of the derivatives 
with respect to the solution constant of the locally homogeneous solutions.
The section $3$ is devoted to the analysis of the nonlinear 
evolution of the multiple slow rolling scalar fields as an 
application of the nonlinear LWL formula constructed in the 
previous section.
In the subsection $3.1$, it is shown that in the slow rolling 
phase the scalar fields evolution equations are simplified
by truncation.
By estimating the truncation error, we establish the accuracy 
of the truncated evolution equations in the slow rolling 
expansion scheme.
In the subsection $3.2$, the ideas of the $\tau$ function 
and the $N$ potential are introduced and they are shown to
enable us to trace the multiple slow rolling scalar fields
in the whole slow rolling phase analytically. 
By adopting the $\tau$ function as the evolution parameter, 
the truncated evolution equations of the multiple slow rolling 
scalar fields  are simplified enough for their solutions 
to be written analytically.
From the scalar fields solutions we can easily calculate the $N$
potential which allows us to calculate the arbitrarily higher order
Bardeen parameter at the end of the slow rolling phase from the 
initial scalar fields perturbations at the first horizon crossing.
In the subsection $3.3$, by the $\tau$ function and the $N$ potential
we calculate the various perturbation variables such as the 
Bardeen parameter, the entropy perturbations, the gravitational
wave perturbation and their spectrum indices in the case of the 
multiple quadratic potential whose truncated evolution can be
exactly solved.
In the subsection $3.4$, we consider the effect of the interaction
between multiple slow rolling scalar fields in the case where the masses
of the scalar fields do not satisfy any resonant relations.
We point out that this problem is related with the well known 
Poincar\'e theorem about the linearization.
In the concrete model, we calculate the various quantities such as 
the $N$ potential and the nonlinear parameters $f_{NL}$, $g_{NL}$
introduced in the paper \cite{Komatsu2001}.
In the subsection $3.5$, we investigate the resonant interaction between 
the multiple slow rolling scalar fields.
We show that the $N$ potential has no singular part by introducing the 
resonant interactions.
The section $4$ is devoted to the discussions.
The appendices are devoted to the proofs of the propositions
presented in the main content of this paper.

\section{Nonlinear LWL formula}

In this section, we derive the nonlinear LWL formula 
in the most general matter composition whose energy momentum tensor 
is divided into $A=(S, f)$ parts where $S$ represents $N_S$ components 
scalar fields $\phi_a$ ($a=1,2, \cdot \cdot \cdot, N_S$) and 
$f$ represents $N_f$ components perfect fluids $\rho_{\alpha}$ 
($\alpha =1,2, \cdot \cdot \cdot, N_f$).
The content of this section is the nonlinear generalization
of the content of the paper \cite{Hamazaki2008}.  
Our results in this section can be applied not 
only to the multiple slow rolling scalar fields, but also
to the multicomponents reheating and the multicomponents 
curvaton decay \cite{Nambu2006} \cite{Hamazaki2008}.

In the subsection $2.1$, we give the evolution equations of the 
locally homogeneous universe in the leading order of the gradient
expansion and we construct all the solutions of these locally 
homogeneous universe evolution equations.
In the subsection $2.2$, we give the definition of the nonlinear gauge
invariant perturbation variables and we give the nonlinear LWL formula
representing the long wavelength limit of the evolutions of the 
gauge invariant perturbation variables in terms of the derivative
with respect to solution constant of the evolutions of the locally
homogeneous universe determined in the subsection $2.1$.

Our theory has two small expansion parameters:
One is $\epsilon$ characterizing the small spatial derivative, 
that is, the small wavenumber,
and the other is $\delta_c$ characterizing the amplitudes of 
the higher order perturbations.
In general, under our present scheme, all the terms are classified 
as $O(\epsilon^k \delta^l_c)$ where $k$, $l$ are appropriate nonnegative 
integers.
Since we are interested in the full nonlinear perturbations,
in the subsection $2.1$ we will not Taylor expand with respect to $\delta_c$.
But since we treat the evolutions of the long wavelength perturbations 
only, we will expand with respect to $\epsilon$ and we will drop 
terms which are small compared to the leading order by $O(\epsilon^2)$
order quantity.
This process corresponds to the leading order of the gradient 
expansion \cite{Rigopoulos2003} \cite{Lyth2005} \cite{Tanaka2007}
and the universe treated in this way is called the locally homogeneous 
universe.
The locally homogeneous evolution equations are very 
similar to the exactly homogeneous evolution equations.
The evolutions of locally homogeneous physical quantities which have
counterparts in the corresponding exactly homogeneous universe 
can be determined as easily as the exactly homogeneous evolutions
are determined.
This is the attractive point of our LWL method.

In our scheme, we can assign 
$\partial_t \tilde{\gamma}_{ij} =O(\delta_c)$, 
$\partial^2_t \tilde{\gamma}_{ij} =O(\delta_c)$ 
where $\tilde{\gamma}_{ij}$ is the unimodular factor
of the spatial metric defined by (\ref{spatialmetricdecompose}),
since we require only that the universe should be the spatially
flat FRW universe in the background level.
So in the leading order of the gradient ($\epsilon$) expansion 
the terms containing $\partial_t \tilde{\gamma}_{ij}$, 
$\partial^2_t \tilde{\gamma}_{ij}$ cannot be dropped.
But in the papers \cite{Lyth2005} \cite{Tanaka2007}, mainly
in order to avoid the computational complexity, the authors 
assigned  $\partial_t \tilde{\gamma}_{ij} =O(\epsilon^2)$, 
$\partial^2_t \tilde{\gamma}_{ij} =O(\epsilon^2)$
and discarded the terms containing $\partial_t \tilde{\gamma}_{ij}$, 
$\partial^2_t \tilde{\gamma}_{ij}$.
But this is confusion between the different small parameters 
$\epsilon$, $\delta_c$ and cannot be justified.
In this section, under the more natural assumption 
$\tilde{\gamma}_{ij} = \delta_{ij} + O(\delta_c)$, we show 
that the evolution of $\tilde{\gamma}_{ij}$ can be solved analytically
and that the evolution equations of the other dynamical variables 
are simple enough to be solved analytically in spite of
inclusion of the terms containing $\partial_t \tilde{\gamma}_{ij}$, 
$\partial^2_t \tilde{\gamma}_{ij}$.
By doing so, in the leading order of the gradient ($\epsilon$) expansion,
the evolutions of all the modes $2 N_S + 4 N_f + 4$ are obtained
consistently in the universe where $N_S$ scalar fields and $N_f$ 
perfect fluids coexists.
In the papers \cite{Lyth2005} \cite{Tanaka2007}, the evolutions of
the several dynamical variables become higher order $O(\epsilon^2)$,
therefore the initial conditions of such variables are unnaturally
constrained to too small quantities.

\subsection{Evolution of the locally homogeneous universe}

First we present the Einstein equations 
$G_{\mu \nu} = \kappa^2 T_{\mu \nu}$ where $\kappa^2 = 8 \pi G$ 
is the gravitational constant, based on the $3+1$ decomposition
based on the paper \cite{Shibata1999}.
The metric 
\begin{equation}
 d s^2 = g_{\mu \nu} d x^{\mu} d x^{\nu},
\end{equation}
is written by
\begin{eqnarray}
 g_{0 0} &=&  - \alpha^2 + \beta_k \beta^k,\\
 g_{0 i} &=& \beta_i,\\
 g_{i j} &=& \gamma_{i j},
\end{eqnarray}
where $\alpha$ is the lapse and $\beta_i$ is the shift
vector and $\beta^i := \gamma^{i j} \beta_j$.
Greek indices take the values $\mu, \nu =0,1,2,3$
and Latin indices take values $i,j =1,2,3$.
The spatial metric $\gamma_{i j}$ is decomposed as
\begin{equation}
 \gamma_{i j} = a^2 \tilde{\gamma}_{i j}, \quad 
 {\rm det} (\tilde{\gamma}_{i j}) =1,
\label{spatialmetricdecompose}
\end{equation}
where $a$ can be interpreted as the nonlinear generalization
of the scale factor.
The extrinsic curvature of the $t={\rm const}$ hypersurface is 
given by
\begin{eqnarray}
 - K_{i j} &=&
 \alpha \Gamma^{0}_{i j} \notag\\
 &=& \frac{1}{2 \alpha}
\left(
 \dot{\gamma}_{i j} - D_i \beta_j - D_j \beta_i
\right),
\end{eqnarray}
where $D_i$ is the covariant derivative with respect to $\gamma_{i j}$.
The extrinsic curvature is decomposed as
\begin{eqnarray}
 K_{i j} &=& \frac{1}{3} \gamma_{i j} K + a^2 \tilde{A}_{i j},\\
 \gamma^{i j} \tilde{A}_{i j} &=& 0.
\end{eqnarray}
Then we obtain
\begin{equation}
 - K = \frac{1}{\alpha} 
\left(
 3 \frac{\dot{a}}{a} - D_i \beta^i
\right).
\label{extrinsictracepart}
\end{equation}
The energy momentum tensor is given by 
\begin{equation}
 T_{\mu \nu} = (\rho + P) u_{\mu} u_{\nu} + P g_{\mu \nu},
\end{equation}
where $\rho$, $P$, $u_{\mu}$ are the energy density, 
the pressure, and the $4$ velocity of the total system,
respectively.
The $4$ velocity $u^{\mu}$ is written as
\begin{eqnarray}
 u^0 &=& \left[ 
\alpha^2 - 
\left( \beta_k + v_k \right) \left( \beta^k + v^k \right)   
         \right]^{- 1/2},\\
 u^i &=& u^0 v^i,
\end{eqnarray}
where $v_i$ is the $3$ velocity of the total system and
$v^i := \gamma^{i j} v_j$. 
By using $n_{\mu}:=(- \alpha, 0,0,0)$ which is the unit vector
normal to the time slices, the $3+1$ decomposition of the energy
momentum tensor is given by
\begin{eqnarray}
 E &:=& T_{\mu \nu} n^{\mu} n^{\nu} 
     = (\rho + P) \left( \alpha u^0 \right)^2 -P,\\
 J_j &:=& - T_{\mu \nu} n^{\mu} \gamma^{\nu}_{\> j}
       = (\rho + P) \alpha u^0 u_j,\\
 S_{i j} &:=& T_{i j}
           = (\rho + P) \left( u^0 \right)^2  
          \left( \beta_i + v_i \right) \left( \beta_j + v_j \right)      
          + P \gamma_{i j}.
\end{eqnarray}
The Hamiltonian and momentum constraints are written as
\begin{eqnarray}
 R - \tilde{A}_{i j} \tilde{A}^{i j} + \frac{2}{3} K^2 
 &=& 2 \kappa^2 E, \label{hamiltonianconstraint} \\
 D_i \tilde{A}^{i}_{\> j} -\frac{2}{3} D_j K &=&
 \kappa^2 J_j,
\label{momentumconstraint}
\end{eqnarray}
where the indices of $\tilde{A}_{ij}$ is raised by 
$\tilde{\gamma}^{ij}$ which is the inverse matrix of
$\tilde{\gamma}_{ij}$.
The evolution equations for $\gamma_{i j}$ are written as
\begin{eqnarray}
 \left( \partial_t -\beta^k \partial_k \right) a &=& 
 \frac{1}{3} a \left( - \alpha K + \partial_k \beta^k \right)
\label{evolutionscalefactor},\\
 \left( \partial_t -\beta^k \partial_k \right) \tilde{\gamma}_{i j} &=&
 - 2 \alpha \tilde{A}_{i j} 
 + \tilde{\gamma}_{i k} \partial_j \beta^k
 + \tilde{\gamma}_{j k} \partial_i \beta^k
 - \frac{2}{3} \tilde{\gamma}_{i j} \partial_k \beta^k.
\label{evolutionmetric}
\end{eqnarray}
The evolution equations for $K_{i j}$ are given by
\begin{eqnarray}
 \left( \partial_t -\beta^k \partial_k \right) K &=&
 \alpha \left( \tilde{A}_{i j} \tilde{A}^{i j} + \frac{1}{3} K^2
        \right) \notag\\
 && - D_k D^k \alpha + \frac{\kappa^2}{2} 
   \alpha \left( E + S^k_{\> k} \right),
\label{evolutionextrinsic} 
\end{eqnarray}
\begin{eqnarray}
 \left( \partial_t -\beta^k \partial_k \right) \tilde{A}_{i j} &=&
 \frac{1}{a^2} \left[ 
 \alpha \left( R_{i j} - \frac{1}{3} \gamma_{i j} R \right) -
 \left( D_i D_j \alpha - \frac{1}{3} \gamma_{i j} D_k D^k \alpha \right)
               \right] \notag\\
 && + \alpha \left( K \tilde{A}_{i j} 
                 -2 \tilde{A}_{i k} \tilde{A}^k_{\> j} \right)
  + \tilde{A}_{i k} \partial_j \beta^k 
  + \tilde{A}_{j k} \partial_i \beta^k \notag\\
 &&
 - \frac{2}{3} \tilde{A}_{i j} \partial_k \beta^k
 - \frac{\kappa^2 \alpha}{a^2} 
 \left( S_{i j} - \frac{1}{3} \gamma_{i j} S^k_{\> k}
 \right).
\label{evolutionextrinsictraceless}
\end{eqnarray}
$R_{i j}$ is the Ricci tensor of the metric $\gamma_{i j}$ and
$R = \gamma^{i j} R_{i j}$, $S^k_{\> k} = \gamma^{k l} S_{k l}$.

Next under the $3+1$ decomposition, we rewrite the evolution equations
of the energy momentum tensor $T^{\mu}_{A \nu}$ of the $A$ component:
\begin{equation}
 \nabla_{\nu} T^{\nu}_{A \mu} = Q_{A \mu},
\end{equation}
where $A:= (a, \alpha)$, $a$ is the scalar fields index and
$\alpha$ is the perfect fluids index. 
The energy momentum transfer vector $Q_{A \mu}$ is decomposed
as
\begin{eqnarray}
 Q_{A \mu} &=& Q_A u_{\mu}+ f_{A \mu},\\
 u^{\mu} f_{A \mu} &=& 0,\label{momentumtransferdef}
\end{eqnarray}
where $Q_A$, $f_{A \mu}$ are the energy transfer,
the momentum transfer vector of the $A$ component, 
respectively. 
The energy momentum tensor of the total system $T^{\mu}_{\> \nu}$
can be expressed as the sum of the each component energy momentum
tensor $T^{\mu}_{A \nu}$:
\begin{equation}
 T^{\mu}_{\> \nu} = \sum_A  T^{\mu}_{A \nu}
\end{equation}
The energy momentum conservation, that is, the Bianchi identity
gives
\begin{equation}
 \sum_A Q_{A \mu} =0.
\end{equation}
We consider the scalar field component $\phi = (\phi_a)$.
The energy momentum tensor of the scalar fields part is given by
\begin{equation}
 \left( T^{\mu}_{\> \nu} \right)_S = 
 \nabla_{\mu} \phi \cdot \nabla_{\nu} \phi
 - \frac{1}{2} \left[     
 g^{\rho \sigma} \nabla_{\rho} \phi \cdot \nabla_{\sigma} \phi
 + 2 U        \right] g_{\mu \nu}  
\end{equation}
The energy momentum tensor of the scalar fields part $(T_{\mu \nu})_S$
can be written as the perfect fluid form by identifying 
\begin{eqnarray}
 \rho_S &=& -
 \frac{1}{2}
 g^{\rho \sigma} \nabla_{\rho} \phi \cdot \nabla_{\sigma} \phi
 + U,\\
 P_S &=& -
 \frac{1}{2}
 g^{\rho \sigma} \nabla_{\rho} \phi \cdot \nabla_{\sigma} \phi
 - U,\\
 u_{a \mu} &=& -
 \frac{1}
 {{\rm sign} (\dot{\phi}_a) 
 \left(- g^{\rho \sigma} \nabla_{\rho} \phi \cdot \nabla_{\sigma} \phi 
 \right)^{1/2}}
 \partial_{\mu} \phi_a, 
\end{eqnarray}
where the minus sign of $u_{a \mu}$ is adopted by requiring 
$u_{a 0} = - \alpha + O(\epsilon^2)$ in the gradient expansion scheme,
where $\epsilon$ is the small parameter characterizing the spatial
derivative defined below.
When we assume that the energy momentum transfer vector of the scalar fields
part $(Q_{\mu})_S$ is given by
\begin{equation}
 (Q_{\mu})_S = S_a \nabla_{\mu} \phi_a,
\end{equation}
where the source function $S_a$ describes the energy transfer 
from the scalar field $\phi_a$ to other components,
the evolution equation of the scalar fields components 
$\nabla_{\nu} (T^{\nu}_{\> \mu})_S = (Q_{\mu})_S$
holds if the scalar field $\phi_a$ satisfies 
the phenomenological equations of motion
of the scalar field $\phi_a$:
\begin{equation}
 \Box \phi_a - \frac{\partial U}{\partial \phi_a} = S_a,
\label{phenomenologicalphia}
\end{equation}
where 
\begin{eqnarray}
  \Box \phi_a &=& \left( - g \right)^{- 1/2} 
 \partial_{\mu} 
 \left[ 
 \left( - g \right)^{1/2}
 g^{\mu \nu} \partial_{\nu} \phi_a 
 \right],\\ 
  g &:=& {\rm det} (g_{\mu \nu}).
\end{eqnarray}

Since we are interested in the cosmological perturbations on
superhorizon scales, we put the gradient expansion 
assumption defined by
\begin{equation}
 \partial_i = O(\epsilon), \quad
 \beta_i = O(\epsilon), \quad
 v_i = O(\epsilon), \quad
 f_{A i} = O(\epsilon), 
\end{equation}
where $\epsilon$ is the small parameter characterizing 
the small wavenumber of the cosmological perturbations.
By the assumption $\partial_i = O(\epsilon)$, 
we assume that the spatial scale of 
all the inhomogeneities is of the order of $1/ \epsilon$, 
that is, all the physical quantities which are approximately 
homogeneous on each horizon can vary on the superhorizon scales.  
The local homogeneity and isotropy in the horizon guarantee 
that the vector quantities such as $\beta_i$, $v_i$, $f_{A i}$
are of the order of $O(\epsilon)$.
Unlike the papers \cite{Lyth2005} \cite{Tanaka2007} where
$\partial_t \tilde{\gamma}_{ij} =O(\epsilon^2)$ is assumed, 
from the requirement that our locally homogeneous universe should 
be the spatially flat FRW universe in the background level,
we assume that
\begin{equation}
 \tilde{\gamma}_{i j} = \delta_{ij} + O(\delta_c).
\end{equation}
where $\delta_c$ is the small parameter characterizing the higher
order perturbations.
Since we consider the leading order of the gradient ($\epsilon$)
expansion without expanding with respect to $\delta_c$, 
we must keep the terms containing 
$\partial_t \tilde{\gamma}_{ij} =O(\delta_c)$,
$\partial^2_t \tilde{\gamma}_{ij} =O(\delta_c)$
as explained in the beginning of this section.

Under the gradient ($\epsilon$) expansion scheme, 
the relations between the total system quantities and 
the each component quantities are given by 
\begin{eqnarray}
 \rho &=& \sum_A \rho_A,\\
 P &=& \sum_A P_A,\\
 h &=& \sum_A h_A,\\
 h v_i &=& \sum_A h_A v_{A i} + O(\epsilon^3),\\
 0 &=& \sum_A Q_A,\\
 0 &=& \sum_A f_{A i},
\end{eqnarray}
where $h_A$ is the $A$ component enthalpy defined by 
$h_A := \rho_A + P_A$.

As for the fluid component $\alpha$, 
$\nabla_{\mu} T^{\mu}_{\alpha 0} = Q_{\alpha 0}$ gives 
\begin{equation}
 \dot{\rho}_{\alpha} = - 3 H (\rho_{\alpha} + P_{\alpha})
 + Q_{\alpha} \alpha + O(\epsilon^2),
\label{energycomponentalpha}
\end{equation}
where $H$ is the Hubble parameter defined by $H := \dot{a}/ a$,
and integrating $\nabla_{\mu} T^{\mu}_{\alpha i} = Q_{\alpha i}$ 
with respect to $t$ gives
\begin{equation}
 h_{\alpha} (\beta_i + v_{\alpha i}) =
 \frac{\alpha}{a^3} C_{\alpha i} +
 \frac{\alpha}{a^3} \int_{t_0} dt \alpha a^3
 \left[
- \partial_i P_{\alpha} - \frac{1}{\alpha} D_i \alpha h_{\alpha} 
+ Q_{\alpha i}
 \right]
+ O(\epsilon^3),
\label{integratevelocityalpha}
\end{equation}
where $t_0$ is the initial time and 
$C_{\alpha i} := C_{\alpha i} (\bm{x})$ are the integration constants.

As for the scalar field components,
from (\ref{phenomenologicalphia}),
the equation of motion of the scalar field $\phi_a$ is given by
\begin{equation}
 \frac{1}{\alpha^2} 
\left[
 \ddot{\phi}_a + 3 \frac{\dot{a}}{a} \dot{\phi}_a
 - \frac{\dot{\alpha}}{\alpha} \dot{\phi}_a
\right]
  + \frac{\partial U}{\partial \phi_a} + S_a
 = O(\epsilon^2).
\label{localhomoscalar}
\end{equation}
Since the source function $S_a$ is the scalar quantity,
we can assume that $S_a$ is the function of other scalar quantities.
As such scalar quantities, we can adopt $\phi_a$,
\begin{equation}
 T_{2 a} := {\rm sgn} (\partial_0 \phi_a)
 \left(- g^{\rho \sigma} \nabla_{\rho} \phi_a \nabla_{\sigma} \phi_a 
 \right)^{1/2}
 = \frac{1}{\alpha} \dot{\phi}_a + O(\epsilon^2),
\end{equation}
or
\begin{equation}
 T_{2 a} := u^{\mu} \nabla_{\mu} \phi_a
 = \frac{1}{\alpha} \dot{\phi}_a + O(\epsilon^2)
\end{equation}
where $u_{\mu}$ is the arbitrary unit time like vector field and 
\begin{equation}
 T_3 := \nabla_{\mu} u^{\mu} = 3 \frac{1}{\alpha}
 \frac{\dot{a}}{a} + O(\epsilon^2).
\end{equation}
Therefore in the leading order of the $\epsilon$ expansion,
the form of the source functions $S_a$ can be given by

\begin{equation}
 S_a = S_a \left( \phi_a, \frac{1}{\alpha} \dot{\phi}_a,
 3 \frac{1}{\alpha} \frac{\dot{a}}{a} \right) 
 + O(\epsilon^2).
\end{equation}
For example,
\begin{equation}
 S_a = \Gamma_a \frac{1}{\alpha} \dot{\phi}_a,
\end{equation}
where $\Gamma_a$ is the decay constant of the scalar 
field $\phi_a$, is the most simple source function.

Since until now we have presented the leading order of the 
gradient ($\epsilon$) expansion of all the locally homogeneous 
evolution equations, we will construct all the solutions of 
the evolutions of the locally homogeneous universe.
From (\ref{evolutionmetric}), we obtain
\begin{equation}
 \tilde{A}_{i j} = - \frac{1}{2 \alpha} 
 \frac{d}{dt} \tilde{\gamma}_{i j} 
 + O(\epsilon^2).
\label{extrinsictracelessdef}
\end{equation}
By substituting the above equation into 
(\ref{evolutionextrinsictraceless}), we obtain
\begin{equation}
 \frac{d^2}{d t^2} M =
 \left( \frac{d}{dt} \ln{\frac{\alpha}{a^3}} \right) \dot{M}
 + \dot{M} M^{-1} \dot{M} + O(\epsilon^2),
\end{equation}
where $M := (\tilde{\gamma}_{i j})$.
By neglecting $O(\epsilon^2)$ order terms,
we obtain the solution as
\begin{equation}
 M = R \exp \left[ \int_{t_0} dt \frac{\alpha}{a^3} T \right],
\label{correctionsolution}
\end{equation}
where $R = R(\bm{x})$, $T = T(\bm{x})$ are time independent 
matrices. 
Since $M$ is a unimodular symmetric matrix for an arbitrary $t$, 
$R$ is unimodular symmetric, $T$ is traceless, and $RT$ is symmetric.
By using (\ref{correctionsolution}), we obtain
\begin{eqnarray}
 \tilde{A}_{i j} \tilde{A}^{i j} &=& \frac{1}{4 \alpha^2}
 {\rm tr} \left( \dot{M} M^{-1} \dot{M} M^{-1} \right)
 = \frac{c_T}{a^6},\\
 c_T &:=& \frac{1}{4} {\rm tr} \left( T^2 \right),
\end{eqnarray}
where $c_T = c_T (\bm{x})$ is a time independent constant.
By using the above results, (\ref{hamiltonianconstraint})
gives
\begin{equation}
 H^2 = \alpha^2 
 \left( \frac{\kappa^2}{3} \rho + \frac{1}{6} \frac{c_T}{a^6}  \right),
\label{modifiedhubble}
\end{equation}
(\ref{evolutionextrinsic}) gives 
\begin{equation}
 \dot{H} = \frac{\dot{\alpha}}{\alpha} H 
 - \frac{\alpha^2}{2} \frac{c_T}{a^6}
 - \frac{\alpha^2 \kappa^2}{2} (\rho + P).
\label{modifiedhdot}
\end{equation}
By eliminating $H$ in (\ref{modifiedhubble}) and 
(\ref{modifiedhdot}), we obtain
the well known continuity equation of the total system 
in the expanding universe as
\begin{equation}
 \dot{\rho} = - 3 H (\rho + P).
\end{equation}
Eqs.(\ref{energycomponentalpha})(\ref{localhomoscalar}) and
the above three evolution equations agree with the exactly 
homogeneous evolution equations if the proper time slicing 
$\alpha =1$ \cite{Kodama1984} and $c_T = 0$.
But we cannot assume $c_T = 0$ since the solution 
constants must satisfy the constraint originating from the momentum 
constraint (\ref{momentumconstraint}) as
\begin{equation}
 \kappa^2 \sum_{\alpha} C_{j \alpha} + \frac{1}{2}
 \partial_i T^i_{\> j} - \frac{1}{4} {\rm tr} 
 \left[ R^{-1} \partial_j R T \right] 
 - \left[ 
 2 a^3 \partial_j \left( \frac{H}{\alpha} \right)
 + \kappa^2 \frac{a^3}{\alpha} \sum_a
 \dot{\phi}_a \partial_j \phi_a  
   \right]_{t_0} = 0.
\label{constantconstraint}
\end{equation}
\textit{For the derivation of (\ref{constantconstraint}), 
see Appendix $A$.}

We consider the long wavelength limit of all the solutions
of the evolution equations of the locally homogeneous universe.
The unimodular factor of the spatial metric $\tilde{\gamma}_{ij}$ 
is obtained by (\ref{correctionsolution}).
These $\tilde{\gamma}_{ij}$ induce the deviation between the 
true locally homogeneous universe and the corresponding  
exactly homogeneous universe, for example 
$c_T$ terms in (\ref{modifiedhubble}) and (\ref{modifiedhdot}).
They contain the scalar adiabatic decaying mode which was 
carefully treated in the papers \cite{Kodama1998} \cite{Hamazaki2008},
since in the scalar part in the linear cosmological perturbation theory 
it induces the deviation between the long wavelength limit
of the true perturbation solutions and the derivative of the exactly
homogeneous universe with respect to the solution constant.
According to the philosophy of the KH construction 
\cite{Kodama1998}, the expressions of $\rho_{\alpha}$, $\phi_a$
related with the exactly homogeneous quantities are obtained 
by solving the evolution equations   
(\ref{energycomponentalpha})
(\ref{localhomoscalar}) under the Hamiltonian 
constraint (\ref{modifiedhubble}) and
the proper time slicing $\alpha = 1$.  
Under the proper time slicing $\alpha = 1$,
(\ref{energycomponentalpha}), (\ref{localhomoscalar}) and
(\ref{modifiedhubble}) are almost the same as 
the counterparts of the exactly homogeneous universe.
Therefore solving the former true locally homogeneous evolution
equations requires as little labor as solving the latter exactly
homogeneous evolution equations.  
As for the velocity perturbations of the fluid
components $h_{\alpha} (\beta_i + v_{\alpha i})$ which are 
the vector quantities not related with the exactly homogeneous 
quantities, their evolutions are given by 
(\ref{integratevelocityalpha}) in whose right hand side
the second integral term contains $P_{\alpha}$, $h_{\alpha}$, 
$\alpha$ which have already been determined by 
the previous process.
The solution constants must satisfy the momentum
constraint (\ref{constantconstraint}).

Let us count the degrees of freedom.
As for (\ref{correctionsolution}), $R$ is unimodular
symmetric, $T$ is traceless, and $RT$ is symmetric, therefore
$R$, $T$ have $5$ degrees of freedom, respectively.
But by the coordinate transformation $\bar{x}^i = f^i (x)$
($i=1,2,3$) the $3$ degrees of freedom of $R$ can be made 
vanishing.   
According to (\ref{energycomponentalpha}) (\ref{localhomoscalar}),
the densities $\rho_{\alpha}$ and the scalar fields $\phi_a$
have $N_f$, $2 N_S$ degrees of freedom, respectively.
According to (\ref{integratevelocityalpha}), the fluids velocities
$v_{\alpha i}$ have $3 N_f$ degrees of freedom.
The momentum constraint (\ref{constantconstraint}) gives $3$ 
constraints.
Therefore the total degrees of freedom is 
$5+5-3+N_f+2 N_S + 3 N_f-3=4+4 N_f+ 2 N_S $.
Then we have obtained all the solutions of the evolution equations of
the locally homogeneous universe in the leading order of the 
gradient ($\epsilon$) expansion.

From the time dependence of all the solutions, we can interpret the 
physical roles of all the solutions.
$2$ from $R$ can be interpreted as the gravitational wave
growing modes.
$5$ from $T$ can be interpreted as the $2$ gravitational wave decaying 
modes, the $1$ adiabatic scalar decaying mode and the $2$ adiabatic vector 
decaying modes.
By the $3$ momentum constraints (\ref{constantconstraint}), $3$
of $C_{j \alpha}$ are adjusted. 
In the remaining $(3 N_f -3)$ $C_{j \alpha}$'s, 
the $(2 N_f -2)$ entropic vector decaying modes and the $(N_f -1)$
entropic scalar decaying modes are contained.
The $N_f$ densities $\rho_{\alpha}$ have the $N_f$ scalar growing
modes and the $N_S$ scalar fields $\phi_a$ have the $N_S$ scalar
fields growing modes and the $N_S$ scalar fields decaying modes.

\subsection{Gauge invariant variables 
and the derivation of the LWL formula}

In this subsection, we give the definitions of
the gauge invariant perturbation variables in the 
arbitrary higher order perturbation theory in the leading order of 
the gradient ($\epsilon$) expansion 
and the LWL formula representing the evolutions 
of these gauge invariant perturbation variables in terms 
of derivative with respect to the solution constant of
the corresponding physical quantities of the locally homogeneous
universe.

As for $M = (\tilde{\gamma}_{i j})$, 
since the homogeneous parts of $R$, $T$ 
defined by (\ref{correctionsolution})
are determined by the fact that the background metric is 
the spatially flat FRW universe,
$R$, $T$  are expanded as
\begin{eqnarray}
 R_{i j} (\bm{x}) &=& \delta_{i j} 
 + \sum_{k=1}^{\infty} \frac{1}{k !}
 \delta^k R_{i j} (\bm{x}),\\
 T_{i j} (\bm{x}) &=& \sum_{k=1}^{\infty} \frac{1}{k !}
 \delta^k T_{i j} (\bm{x}),
\end{eqnarray}
where $\delta$ in the above is the operator generating the 
higher order perturbation quantities.

As shown in the previous subsection, in the leading order
of the gradient ($\epsilon$) expansion, the evolution equations
of the locally homogeneous universe which have counterparts
in the evolution equations of the exactly homogeneous universe
do not contain the spatial derivative.
Therefore in the expression of the solution of the evolution of
an arbitrary locally homogeneous physical quantity (related with 
the exactly homogeneous quantity) $A$ such as
the scalar fields $\phi_a$ and the fluid energy densities $\rho_{\alpha}$, 
all the dependences on the spatial coordinate $\bm{x}$ are contained 
in the (time independent) spatial dependent integration constants 
$C(\bm{x})$:
\begin{equation}
 A = A(t, C(\bm{x})).
\label{AtCspatial}
\end{equation}
Since the solutions of the evolution equations of the locally 
homogeneous universe contain the nonlinear effect in the full
order, the locally homogeneous physical quantity $A$ can be expanded
as
\begin{equation}
 A (t, \bm{x}) = A(t) + \sum_{k=1}^{\infty} \frac{1}{k !}
 \delta^k A (t, \bm{x}).
\label{Aexpanddelta}
\end{equation}
The higher order perturbation effects are induced by the 
dependences on the spatial coordinate $\bm{x}$.
Therefore the each solution constant $C_i (\bm{x})$ can be expanded as
\begin{equation}
 C_i (\bm{x}) = C_i + \sum_{k=1}^{\infty} \frac{1}{k !}
 \delta^k C_i (\bm{x}),
\label{constexpanddelta}
\end{equation}
whose background part is spatially independent.
The $k$-th order nonlinear perturbation $\delta^k A$
can be expressed in terms of the nonlinear perturbations
of the spatial dependent integration constants $\delta^l C(\bm{x})$. 
In this paper, the expressions representing $\delta^k A$ in terms of
$\delta^l C(\bm{x})$ are called the LWL formula.
In order to derive the LWL formula,
we propose a simple mathematical trick.
In the leading order of the gradient ($\epsilon$) expansion,
the expression of the solution of an arbitrary locally homogeneous 
physical quantity $A$ can be written as function of the time $t$ 
and the integration constants $C$ as shown in (\ref{AtCspatial}).
We assume that all the integration constants $C$ depend on 
one parameter $\lambda$ imaginarily instead of $\bm{x}$.
As understood below, $\lambda$ represents the spatial coordinate
dependence symbolically. 
The physical quantity $A$ can be expanded as
\begin{equation}
 A (\lambda) = \sum_{k=0}^{\infty} \frac{1}{k !}
 \lambda^k
 \frac{d^k}{d \lambda^k} A (\lambda) \Bigg|_{\lambda = 0},
\label{Alambdaexpand}
\end{equation}
The full order nonlinear solution is formally recovered by 
setting $\lambda = 1$.
Since both the perturbation $\delta$ and the $\lambda$ differentiation
$d / d \lambda$ are the derivative operators satisfying the same
chain and product rules, and the $k$-th order $\lambda$ differentiation
$d^k / d \lambda^k \cdot \cdot \cdot |_{\lambda =0}$ is multiplied by 
$\lambda^k$, we can use the $\lambda$ differentiation to track the 
algebraic behaviors of the perturbation $\delta$. 
By comparing (\ref{Alambdaexpand}) with (\ref{Aexpanddelta}), 
we can read the correspondences given by
\begin{equation}
 \frac{d^k}{d \lambda^k} A (\lambda) \Bigg|_{\lambda = 0}
 \leftrightarrow
 \delta^k A (t, \bm{x}),
\label{corresA}
\end{equation}

The gauge transformation can be expressed by the Lie derivative
$L(T)$ as 
\begin{equation}
 A(\lambda, \mu) = \exp \{ \mu L(T) \} A(\lambda, \mu=0),
\label{masterofgaugetrans}
\end{equation}
where $\mu$ is the parameter characterizing the size
of the gauge transformation and
$A(\lambda, \mu=1)$ is the transformed variable and 
$A(\lambda, \mu=0)=A(\lambda)$ is the original variable.
When the vector field $T:= T^{\mu} \partial_{\mu}$ 
in the Lie derivative $L(T)$ can be expanded in terms of
$\lambda$, the zeroth order term $T(\lambda =0)$ is zero 
since we consider the infinitesimal gauge transformation.
By differentiating (\ref{masterofgaugetrans}) with respect to $\lambda$,
afterward putting $\lambda = 0$, we obtain the well known
expressions of the gauge transformations of $\delta^k A$
\cite{Bruni1997}:
\begin{eqnarray}
 \tilde{\delta A} &=& L_1 A + \delta A,\\
 \tilde{\delta^2 A} &=& L_2 A + L_1 L_1 A + 2 L_1 \delta A
 +  \delta^2 A,
\end{eqnarray}
where $\tilde{\delta^n A}$ is the gauge transformed 
perturbation variables of $\delta^n A$, and
$L_k$ is the Lie derivative induced by $\delta^k T$;
$L_k := L (\delta^k T)$.
The gauge transformation law (\ref{masterofgaugetrans}) is the 
solution of the differential equation:
\begin{equation}
 \frac{d}{d \mu} A(\lambda, \mu) = L(T) A(\lambda, \mu),
\label{gaugetransdiffeq}
\end{equation}
which is much simpler than the individual gauge transformation
expressions of $\delta^k A$.
As for the vector field $T := T^{\mu} \partial_{\mu}$ 
inducing the Lie derivative $L(T)$, we can assume that
\begin{equation}
 T^i = O(\epsilon),
\label{ourscheme}
\end{equation}
since we consider only the gauge transformations keeping 
the local homogeneity and isotropy in the horizon.

By using the differential equation (\ref{gaugetransdiffeq}), we can 
show the following proposition:

\paragraph{Proposition $1$}

\textit{ For an arbitrary scalar quantity $A$, the perturbation 
quantities $D^n A$, $D^n (\dot{A} / \alpha)$ where $D$ is 
defined by}
\begin{equation}
 D := \frac{d}{d \lambda} - 
 \frac{d a}{d \lambda} \frac{1}{\dot{a}}  
 \frac{d}{d t},
\label{Doperation}
\end{equation}
\textit{are gauge invariant up to order $O(\epsilon^2)$
error.
For the lapse function $\alpha$, ${\cal A}_n$ defined by}
\begin{equation}
 {\cal A}_n :=
\left( \frac{d}{d \lambda} - 
 \frac{d}{d t} \frac{1}{\dot{a}} 
 \frac{d a}{d \lambda}
              \right)^n
 \alpha.
\label{calAN}
\end{equation}
\textit{is gauge invariant up to order $O(\epsilon^2)$ error.}

\textit{For the proof, see Appendix $C$.}

$D^n A$, $D^n (\dot{A} / \alpha)$ and ${\cal A}_n$ are higher order
generalization of $D A$, $D \dot{A}$ and ${\cal A}$ defined in the paper
\cite{Hamazaki2008}, respectively.
$D^n (\dot{A} / \alpha)$ and ${\cal A}_n$ are not independent.
For example
\begin{equation}
  D \left( \frac{\dot{A}}{\alpha} \right) = 
 \frac{1}{\alpha^2} \left\{ 
 (D A)^{\cdot} \alpha - \dot{A} {\cal A}_1
 \right\}.
\end{equation}
Therefore by putting $A=\phi_a$, $A=\rho_{\alpha}$,
we can use $D^n \phi_a$, $D^n (\dot{\phi}_a / \alpha)$, 
$D^n \rho_{\alpha}$ as the independent perturbation variables.
Our $D A$ is almost the same as the well known gauge invariant 
perturbation variable, for example, for $A=\phi_a$ 
the Sasaki Mukhanov variable in the linear perturbation theory 
\cite{Kodama1984} \cite{Kodama1987}
\cite{Mukhanov1992} \cite{Hamazaki2008}.
Our $D^2 A$ agrees almost perfectly with the gauge invariant quantities
introduced by the second order gauge invariant perturbation theory
\cite{Malik2004} \cite{Malik2005} \cite{Nakamura2006}.
The reason of the tiny deviation between our gauge invariant
perturbation quantity and the gauge invariant perturbation 
quantity of the first and the second order
gauge invariant perturbation theory is that we assume 
$T^i =O(\epsilon)$ based on the gradient expansion scheme.
In our scheme (\ref{ourscheme}), we can regard the scale factor
$a$ as the scalar quantity up to $O(\epsilon^2)$.
Therefore in the same way as in the proof of Proposition $1$
we can show that 
\begin{eqnarray}
 \zeta_n &:=& \bar{\delta}^n \ln{a},
\label{generalbardeenparam}\\
\bar{\delta} &:=&
\frac{d}{d \lambda} - 
 \frac{d \rho}{d \lambda} \frac{1}{\dot{\rho}} \frac{d}{d t}
\label{deltabarop}
\end{eqnarray}
are gauge invariant up to $O(\epsilon^2)$.
$\zeta_n$ is the higher order generalization of the 
well known Bardeen parameter \cite{Bardeen1980} \cite{Wands2000}.

We discuss the influence of the replacement of the evolution
parameter.
Concretely we consider replacing the old evolution parameter $t$
with the new evolution parameter as the scale factor $a$.
With the scale factor $a$ as the evolution parameter,
the $D$ operation defined by (\ref{Doperation}) can be written
in the more simple form. 
$d / d \lambda$ in (\ref{Doperation}) is the $\lambda$ derivative  
with $t$ fixed, that is, operating on the locally homogeneous 
physical quantity such as $\rho_{\alpha}$, $\phi_a$
expressed by using $t$ as the evolution parameter.
However an arbitrary locally homogeneous quantities can be also expressed by
using the scale factor $a$ as the evolution parameter
instead of $t$.
In this case, we can consider the differentiation 
$(\partial / \partial \lambda)_a$ taken at fixed $a$.
The differential operators in the two group
$(d / d \lambda, d / d t)$ 
$( (\partial / \partial \lambda)_a , (\partial / \partial a)_a )$ 
are commutative in the each group, but the differential 
operators belonging to the different groups, 
for example $d / d \lambda$ and $(\partial /\partial a)_a$,
are not commutative.
By using the relation 
\begin{equation}
 \frac{d}{d \lambda} = \left(\frac{\partial}{\partial \lambda} \right)_a+
 \frac{d a}{d \lambda} \left(\frac{\partial}{\partial a} \right)_a,
\end{equation}
the $D$ operation (\ref{Doperation}) can be expressed 
much more simply as
\begin{equation}
 D = \left( \frac{\partial}{\partial \lambda} \right)_a.
\end{equation}
Therefore $D^n A$, $D^n (\dot{A} / \alpha)$ can be written 
in the very simple way as
\begin{equation}
 D^n A = \left( \frac{\partial^n}{\partial \lambda^n} \right)_a A, \quad \quad
 D^n \left( \frac{\dot{A}}{\alpha} \right) =
\left( \frac{\partial^n}{\partial \lambda^n} \right)_a
\left( \frac{\dot{A}}{\alpha} \right).
\end{equation}
In this way, in the LWL formalism where the scale
factor $a$, not $t$, is used as the evolution parameter
\cite{Hamazaki2002} \cite{Hamazaki2004} \cite{Hamazaki2008},
the expressions of the solutions of $D^n A$, $D^n (\dot{A} / \alpha)$
can be obtained by calculating only a single term written with
the higher order $(\partial / \partial \lambda)_a$ derivative of 
the solution of the corresponding locally homogeneous physical
quantity.
In the same way, $\bar{\delta}$ defined by (\ref{deltabarop})  
can be expressed in the word of the scale factor $a$ as
\begin{equation}
 \bar{\delta} =
\left( \frac{\partial}{\partial \lambda} \right)_a
- \left\{ \left(\frac{\partial \rho}{\partial \lambda} \right)_a
 \Big/ \left(\frac{\partial \rho}{\partial a} \right)_a \right\}
 \left( \frac{\partial}{\partial a} \right)_a.
\label{deltabarscalefactor}
\end{equation}
As for the Bardeen parameter defined by 
(\ref{generalbardeenparam}), we can show the following 
propositions.

\paragraph{Proposition $2$}

\textit{If $\zeta_1$ is conserved for arbitrary values of
integration constants $C(\lambda = 0)$, 
all $\zeta_n$ ($n \ge 2$) are also
conserved, and they can be expressed as}
\begin{equation}
 \zeta_n = \left( \frac{\partial}{\partial \lambda} \right)^{n-1}_a
 \zeta_1
\end{equation}

\textit{For the proof, see Appendix $D$}
 
\paragraph{Proposition $3$}

\textit{When $P = P(\rho)$ holds, $\zeta_n$ ($n \ge 1$) can be 
expressed as}
\begin{equation}
 \zeta_n = \frac{1}{3} 
\left( \frac{\partial^n}{\partial \lambda^n} \right)_a
\left[ \int d \rho \frac{1}{\rho + P(\rho)} \right]
\end{equation}
\textit{and all $\zeta_n$ ($n \ge 1$) are conserved.}

\textit{For the proof, see Appendix $E$}

All the definitions of the nonlinear gauge invariant perturbation
variables in this subsection are written in terms of the $\lambda$
differentiations.
As explained at the beginning of this subsection, the $\lambda$
differentiation implies not only the symbol of the higher order 
nonlinear perturbation of the physical quantity, but also the process 
of taking the derivative with respect to the integration constants
of the corresponding locally homogeneous quantity.
Therefore all the definitions of the nonlinear gauge invariant
perturbation variables in terms of the $\lambda$ differentiations
can also be regarded as the LWL formulae themselves, so from now on
they will be called the LWL formulae.

\section{Nonlinear evolution of 
the multiple slow rolling scalar fields}

In this section, as the application of the LWL formula
derived in the previous section, we consider the evolutions
of the long wavelength nonlinear cosmological perturbations in the 
universe dominated by the multiple slow rolling scalar 
fields.
The $\tau$ function and the $N$ potential introduced in this 
section are the useful tools for tracing analytically
the evolutions of the multiple slow rolling scalar fields 
in the long time interval.
We calculate spectral indices of the linear cosmological perturbations.
In the interacting system, we derive the formulae giving the amplitudes
of the nonlinear Bardeen parameters at the end of the slow rolling phase
in terms of the initial scalar fields perturbations, and calculate the 
nonlinear parameters $f_{NL}$, $g_{NL}$ \cite{Komatsu2001} representing
the non-Gaussianity of the Bardeen parameter.

\subsection{Evolution of the multiple slow rolling 
scalar fields}

In the slow rolling phase, the scalar fields $\phi_a$ roll
slowly on the potential $U$.
The potential energy $U$ which hardly changes triggers 
the exponential expansion of the universe.
The Hubble parameter is large compared to the masses of the 
scalar fields.
The ratio of the kinetic energy part in the whole energy 
density $\rho$ is small compared to the contribution from
the potential energy $U$.
In the investigations of the evolutions of the scalar fields
under this situation, it is effective to use the transformation by
which the evolution equations of the slow rolling scalar fields
system are greatly simplified, that is the effects of the time 
derivatives of the scalar fields $p_a := \dot{\phi}_a / \alpha$
on the evolutions of the scalar fields $\phi_a$ are eliminated.

In this section, we consider the multiple slow rolling 
scalar fields $\phi_a$ under the conditions:
\begin{equation}
 S_a = 0, \quad \quad
 U = \sum_a \frac{1}{2} m^2_a \phi^2_a + U_{{\rm int}}.
\label{chaoticpotential}
\end{equation}
where $U_{{\rm int}}$ is the sum of $m$-th order monomials ($m \ge 3$).
As the independent variables, we adopt $\phi_a$ and
$p_a := \dot{\phi}_a / \alpha$.
By nondimensionalizing the dynamical variables as
\begin{equation}
 \frac{a}{a_0} \to a, \quad
 \frac{\phi_a}{\phi_0} \to \phi_a, \quad
 \frac{p_a}{p_0} \to p_a, \quad
 \frac{c_T}{c_{T 0}} \to c_T,
\end{equation}
and the parameter as
\begin{equation}
 \frac{m_a}{m_0} \to m_a, 
\end{equation}
that is
\begin{equation}
 \frac{\rho}{m^2_0 \phi^2_0} \to \rho,
\end{equation}
we obtain the dimensionless parameters:
\begin{equation}
 \epsilon_{\ast} := \frac{\sqrt{3}}{\kappa \phi_0}, \quad
 \eta^2 := \frac{p^2_0}{m^2_0 \phi^2_0}, \quad  
 \nu^2 := \frac{1}{2} \frac{c_{T 0}}{\kappa^2 m^2_0 \phi^2_0}
 \frac{1}{a^6_0}.
\end{equation}
$\epsilon_{\ast}$ is the small constant representing
the ratio of the mass scale to the Hubble parameter. 
This $\epsilon_{\ast}$ is different from  
$\epsilon$ characterizing the small wavenumber in the previous section.
$\eta^2$ is positive constant since we consider only the scalar fields
with positve definite kinetic parts.
We assume that $\epsilon_{\ast} \sim \eta \sim \nu \ll 1$.
Then the evolution equations of $\phi_a$, $p_a$ and $c_T$ are given by
\begin{eqnarray}
 \frac{d}{d N} \phi_a &=&
 \frac{\epsilon_{\ast}}{\eta} \frac{1}{\rho^{1/2}} \eta^2 p_a,
\label{evolutionphi}\\
 \frac{d}{d N} p_a &=& - 3 p_a -
 \frac{\epsilon_{\ast}}{\eta} \frac{1}{\rho^{1/2}} 
 \frac{\partial U}{\partial \phi_a},
\label{evolutionp}\\
 c_T &=& {\rm const},
\end{eqnarray}
where the evolution parameter is the e-folding number 
of the scale factor $N := \ln{a}$ and the energy density $\rho$
is given by
\begin{equation}
 \rho = \frac{\eta^2}{2} \sum_a p^2_a + U + \nu^2 c_T \frac{1}{a^6}.
\end{equation}
By using $p^{(1)}_a$ defined by
\begin{equation}
p^{(1)}_a := p_a + \frac{1}{3}
 \frac{\epsilon_{\ast}}{\eta} \frac{1}{\rho^{1/2}} 
 \frac{\partial U}{\partial \phi_a},
\end{equation}
which represents the deviation of $p_a$ from the truncated slow rolling 
solution,
the evolution equations (\ref{evolutionphi})(\ref{evolutionp})
can be rewritten as
\begin{eqnarray}
 \frac{d}{d N} \phi_a &=&
 \eta^2 F_a (\phi) + \eta^2 f_a (\phi, p^{(1)}, c_T, N),
\label{evolutionnewphi}\\
 \frac{d}{d N} p^{(1)}_a &=& - 3 p^{(1)}_a 
 + \eta^2 g_a (\phi, p^{(1)}, c_T, N)
\label{evolutionnewp}
\end{eqnarray}
where 
\begin{equation}
 F_a (\phi) := - \frac{1}{3} 
 \frac{\epsilon^2_{\ast}}{\eta^2} \frac{1}{U} 
 \frac{\partial U}{\partial \phi_a},
\end{equation}
and 
\begin{equation}
 |f_a| \le |p^{(1)}| + \eta^2, \quad \quad
 |g_a| \le 1,
\end{equation}
for an appropriate complex domain containing the 
real interval where we consider the motion of 
$\phi_a$, $p^{(1)}_a$, $N$.  
In this section, in all the inequalities we omit all the finite
constants and $|p^{(1)}|$ is interpreted as the quantity 
bounded by $M |p^{(1)}|$ for some positive constant $M$ and
\begin{equation}
 |p^{(1)}| := \sum_a |p^{(1)}_a|. 
\end{equation}
From now on, we will simply write $p^{(1)}_a$ as $p_a$  
for notational simplicity.
In this section, we consider the evolution for 
$0 \le N \le 1 / \eta^2$ during which the scalar fields roll
slowly on the potential $U$.
For a function $f(N)$, let us define $\| f \|$ by   
\begin{equation}
 \| f \| := \sup_{0 \le N \le 1 / \eta^2} |f(N)|.
\end{equation}
Under these notations, following propositions hold.

\paragraph{Proposition $4$}

\textit{ Let $k$ nonnegative integer, and $\delta_c$ small positive
constant. Under the initial conditions}
\begin{equation}
 \frac{\partial^k}{\partial \lambda^k} \phi(0), \quad
 \frac{\partial^k}{\partial \lambda^k} p(0), \quad
 \frac{\partial^k}{\partial \lambda^k} c_T \sim \delta^{k}_c,
\label{initialpartialk}
\end{equation}
\textit{for $0 \le N \le 1 / \eta^2$, the upper bounds of the
independent variables are given by}
\begin{eqnarray}
 \left| \left( \frac{\partial^k}{\partial \lambda^k} \right)_a \phi 
 \right| &\le&
 \delta^k_c,
\label{partialkphi}\\
 \left| \left( \frac{\partial^k}{\partial \lambda^k} \right)_a p 
 \right| &\le&
 \delta^k_c ( e^{- 3 N} + \eta^2),
\label{partialkp}
\end{eqnarray}

\textit{For the proof, see Appendix $F$.}

In the above and from now on, as for $\lambda$ differentiations 
of the physical quantities at the initial time $N=0$, 
the suffixes $a$ implying ``$a$ fixed'' are omitted since 
what the $\lambda$ differentiations operate on do not contain 
any $a = e^N$ dependent parts.

\paragraph{Proposition $5$}

\textit{Let $k$ nonnegative integer.
The differences $\Delta \phi_a := \phi_a - \bar{\phi}_a$ where $\phi$
obey the exact evolution equations (\ref{evolutionnewphi})
(\ref{evolutionnewp}) and $\bar{\phi}$ obey the truncated evolution
equations as}
\begin{equation}
 \frac{d}{d N} \bar{\phi} = \eta^2 F(\bar{\phi}),
\label{truncatedphibar}
\end{equation}
\textit{are bounded as} 
\begin{equation}
 \left\| \left( \frac{\partial^k}{\partial \lambda^k} \right)_a 
 \Delta \phi \right\|
 \le \eta^2 \delta^k_c \sim \eta^2
 \left\| \left( \frac{\partial^k}{\partial \lambda^k} \right)_a 
 \phi \right\|,
\label{partialkerror}
\end{equation}
\textit{under the initial conditions }
\begin{equation}
 \frac{\partial^k}{\partial \lambda^k} \Delta \phi (0) = 0.
\label{initialerror}
\end{equation}

\textit{For the proof, see Appendix $G$.}

According to Proposition $5$, if we want to investigate the evolution
of the leading order in the slow rolling ($\eta^2$) expansion, 
we only have to solve rather simple evolution equations
(\ref{truncatedphibar}).
In the following subsections, we study the evolutionary behaviors 
of the above evolution equations (\ref{truncatedphibar}).

\subsection{the $\tau$ function and the $N$ potential}

In this subsection, we introduce the $\tau$ function 
and the $N$ potential which enable us to trace 
the multiple slow rolling scalar fields in the long 
time interval analytically.
By using the $\tau$ function as the evolution parameter,
the evolutions of the scalar fields $\phi_a$ and the 
old evolution parameter $N$ can be expressed in the 
form of the simple analytic function of $\tau$.
From these expressions of $\phi_a$ in terms of the 
$\tau$ function, we can calculate the $N$ potential
in the simple way.
The $N$ potential is written in terms of the initial values
of the scalar fields $\phi_a (0)$ only.
It not only represents the difference between 
the e-folding number of the scale factor at the end of 
the slow rolling phase and that at the first horizon crossing,
but also has the complete information as to the nonlinear
curvature perturbations.
The $\lambda$ differentiations of the $N$ potential generate
the S formulae connecting the amplitudes of all the higher 
order Bardeen parameters at the end of the slow rolling phase
with the scalar fields perturbations at the first horizon crossing.

Under the conditions (\ref{chaoticpotential}),
we will investigate the evolutionary behaviors of the 
solution of the evolution equations which are obtained
by the truncation explained in the previous subsection.
\begin{equation}
 \frac{d}{d N} \phi_a = - \frac{1}{\kappa^2} \frac{1}{U}
 \frac{\partial U}{\partial \phi_a}.
\label{evolutionstart}
\end{equation}
From the present subsections, we will call off the nondimensionalization
in the previous subsection, since the truncated evolution equations
(\ref{evolutionstart}) have already been made sufficiently simple.
In the multiple scalar fields case, the evolution equations 
(\ref{evolutionstart}) cannot be solved analytically, and 
the scalar fields $\phi_a$ cannot be expressed in the form of
the well known function of $N$.
So by replacing the old evolution parameter $N$ with the new
evolution parameter $\tau$, we decompose (\ref{evolutionstart}) 
into two parts:
\begin{eqnarray}
  \frac{d}{d \tau} \phi_a &=& - \frac{\partial U}{\partial \phi_a},
\label{firstevolution}\\
 \frac{d}{d \tau} N &=& \kappa^2 U.
\label{secondevolution}
\end{eqnarray}
The new evolution parameter $\tau$ introduced in the above equations 
will be called the $\tau$ function from now on.
By introducing the $\tau$ function, the evolution equations become
simple enough to be solved analytically. 
In the potential (\ref{chaoticpotential}), we can easily solve 
(\ref{firstevolution}) by iteration, then we can get $\phi_a (\tau)$ 
expressed in the form of the analytic functions of the $\tau$ function.
By substituting these $\phi_a (\tau)$ into $U$ in (\ref{secondevolution}),
and integrating (\ref{secondevolution}) with respect to the $\tau$
function, we obtain the expression of the old evolution parameter $N$
in terms of the new evolution parameter $\tau$:
\begin{equation}
 N = \int^{\tau}_0 d \tau \kappa^2 U.
\label{preNpotential}
\end{equation}
The expressions of $\phi_a$, $N$ in terms of the $\tau$ function
describe the dynamical evolutions of our multiple slow rolling 
scalar fields system completely.

From this subsection, we adopt the $\tau$ function as the evolution 
parameter.
We introduce $(\partial / \partial \lambda)_{\tau}$ as the $\lambda$ 
differentiation taken at the fixed $\tau$, that is, operating on the 
locally homogeneous quantities expressed by using the $\tau$ function
as the evolution parameter.
In order to exaggerate the fact that the $\tau$ derivative and
$(\partial / \partial \lambda)_{\tau}$ are commutative, 
we use $(\partial / \partial \tau)_{\tau}$ 
as the $\tau$ derivative from now on.
By using $(\partial / \partial \lambda)_{\tau}$, 
$(\partial / \partial \tau)_{\tau}$,
$D$ defined by (\ref{Doperation}) can be expressed as
\begin{equation}
 D = \left(\frac{\partial}{\partial \lambda} \right)_{\tau} 
 - \left\{ \left(\frac{\partial a}{\partial \lambda} \right)_{\tau} 
 \Big/ \left( \frac{\partial a}{\partial \tau} \right)_{\tau} \right\}
 \left( \frac{\partial}{\partial \tau} \right)_{\tau},
\label{Doperationtaufunction}
\end{equation}
and $\bar{\delta}$ defined by (\ref{deltabarop})
can be expressed as
\begin{equation}
 \bar{\delta} = \left( \frac{\partial}{\partial \lambda} \right)_{\tau} 
- \left\{ \left( \frac{\partial U}{\partial \lambda} \right)_{\tau} 
\Big/  \left( \frac{\partial U}{\partial \tau} \right)_{\tau} \right\}
 \left( \frac{\partial}{\partial \tau} \right)_{\tau},
\label{deltabaroptaufunction}
\end{equation}
where we used 
\begin{equation}
 \frac{d}{d \lambda} = 
 \left(\frac{\partial}{\partial \lambda} \right)_{\tau}
 + \frac{d \tau}{d \lambda} 
 \left(\frac{\partial}{\partial \tau} \right)_{\tau},
\end{equation}
and $\rho = U$ which holds under the present truncation 
(\ref{evolutionstart}).
Then by using the $\tau$ function as the evolution parameter,
the Bardeen parameter $\zeta_n$ defined by (\ref{generalbardeenparam})
can be decomposed as
\begin{equation}
 \zeta_n = \frac{\partial^n}{\partial \lambda^n} \bar{N}
 - \frac{\kappa^2}{4} \bar{\delta}^n A(0,0).
\label{bardeenparamdecomposition}
\end{equation}
By $\bar{N}$, we represent 
\begin{equation}
 \bar{N} := \int^{\infty}_{0} d \tau \kappa^2 U,
\label{Npotentialdef}
\end{equation}
and this $\bar{N}$ will be called the $N$ potential from now on.
In the above and from now on, as for $\lambda$ differentiations 
of the physical quantities at the initial time $\tau = 0$, 
the suffixes $\tau$ implying ``$\tau$ fixed'' are omitted since 
what the $\lambda$ differentiations operate on do not contain 
any $\tau$ dependent parts.   
$A(2n, k)$ is defined by
\begin{eqnarray}
A(0,0) &:=& 4 \int^{\infty}_{\tau} d \tau U,\\
A(2n,k) &:=& \left( - \frac{1}{2} \right)^n 
\left(
\frac{\partial^n}{\partial \tau^n}
\frac{\partial^k}{\partial \lambda^k} \right)_{\tau}
 A(0,0).
\end{eqnarray}
By using $A(2n, k)$, $\bar{\delta}$ can be expressed as
\begin{equation}
 \bar{\delta} = \left(\frac{\partial}{\partial \lambda} \right)_{\tau}
+ \frac{1}{2} \frac{A(2,1)}{A(4,0)}
\left( \frac{\partial}{\partial \tau} \right)_{\tau}.
\end{equation}

Afterward we will prove the fact that as for the Bardeen parameter
$\zeta_n$ at the end of the slow rolling phase the first term in 
(\ref{bardeenparamdecomposition}) has leading contribution in the 
slow rolling expansion scheme.
In order to prove this statement, we put forth several assumptions.
All the investigations in this section will be established 
under the following assumptions:
\begin{itemize}
 \item[(i)] All the masses of the scalar fields are of the same 
  order.
\begin{equation}
 m^2_a \sim m^2, 
\end{equation}
where
\begin{equation}
 m^2 := \min_a \{ m^2_a \}. 
\end{equation}
 \item[(ii)] The interaction potential $U_{{\rm int}}$ is 
 the sum of the $m$-th order monomials ($m \ge 3$) and 
 satisfies
\begin{equation}
 | U_{{\rm int}} | \le \mu_c m^2 \frac{1}{\phi_0} |\phi|^3,
\end{equation}
 for $|\phi| \le \phi_0$.
 $\phi_0$ is defined by the positive constant of the order of
\begin{equation}
 \phi_0 \sim \phi_a (0),
\end{equation}
 and $|\phi|$ is defined by
\begin{equation}
 |\phi|^2 := \sum_a |\phi_a|^2, 
\end{equation}
 and $\mu_c$ is the small positive constant characterizing 
 the interaction strength.
 \item[(iii)] The $n$-th order perturbation at the initial time
 $\tau = 0$ is of the order of
\begin{equation}
 \frac{\partial^k}{\partial \lambda^k} \phi_a (0)
 \sim \delta^k_c \phi_0,
\end{equation}
 where $\delta_c$ is the small constant characterizing the 
 perturbation size.
\end{itemize}
Under these assumptions, the following proposition holds.

\paragraph{Proposition $6$}

\textit{The following estimations hold.}
\begin{eqnarray}
 \frac{\partial^k}{\partial \lambda^k} \bar{N}
 &=& \frac{\kappa^2}{4} \sum_a
 \frac{\partial^k}{\partial \lambda^k}
 [ \phi^2_a (0) ] + < \kappa^2 \mu_c \delta^k_c \phi^2_0  >,
\label{Prop6result1}\\
 A(2n, k) &=& \sum_a \left( m^2_a \right)^n
 \frac{\partial^k}{\partial \lambda^k}
 [ \phi^2_a (0) ] \exp{[- 2 m^2_a \tau ]}
+ < \mu_c m^{2n} \delta^k_c \phi^2_0 \exp{[- 2 m^2 \tau ]} >,
\label{Prop6result2}
\end{eqnarray}
\textit{where $<M>$ is the quantity bounded by $M$.}

\textit{For the proof, see Appendix $H$.}

At the end of the slow rolling phase when the Hubble
parameter is of the order of the scalar fields mass, 
the values of the scalar fields are of the order of the 
Planck mass $\phi_a (\tau) \sim 1 / \kappa$.
According to Proposition $6$, at the end of the slow rolling 
phase, we obtain the estimations as
\begin{eqnarray}
 \frac{\partial^k}{\partial \lambda^k} \bar{N}
 &\sim& \kappa^2 \delta^k_c \phi^2_0,\\
 A(2n, k) &\sim& m^{2 n} \delta^k_c \phi^2_0
 \left( \frac{m}{H_0} \right)^2, 
\end{eqnarray}
where $H_0$ is the Hubble parameter at the initial time
$\tau = 0$.
By using the above estimations,
the second term of (\ref{bardeenparamdecomposition})
can be estimated as
\begin{equation}
 - \frac{\kappa^2}{4} \bar{\delta}^n A(0, 0)
\sim \kappa^2 \delta^n_c \phi^2_0
 \left( \frac{m}{H_0} \right)^2
\sim 
\frac{\partial^n}{\partial \lambda^n} \bar{N} \cdot
 \left( \frac{m}{H_0} \right)^2.
\end{equation}
Since the second term of (\ref{bardeenparamdecomposition}) 
is suppressed by the slow rolling parameter $(m/ H_0)^2$
compared to the first term, the main part of the nonlinear 
$n$-th order Bardeen parameter $\zeta_n$ 
at the end of the slow rolling phase
is given by the $\lambda$ derivative of the $N$ potential:
\begin{equation}
 \zeta_n = \frac{\partial^n}{\partial \lambda^n} \bar{N}.
\label{Bardeenleading}
\end{equation}
So in order to obtain the final amplitude of the Bardeen parameter
$\zeta_n$, we have only to calculate the $N$ potential $\bar{N}$.

In particular case, the $N$ potential can be calculated very easily.
When the scalar fields potential is written 
in the separable form
\begin{equation}
 U = \sum_a U_a (\phi_a),
\end{equation}
the $N$ potential can also be expressed in the separable form
as 
\begin{equation}
 \bar{N} = \sum_a \bar{N}(a),
\end{equation}
where 
\begin{equation}
 \bar{N}(a) := \kappa^2 \int^{\phi_a (0)}_0 d \phi_a 
 \left( U_a \Big/ \frac{\partial U_a}{\partial \phi_a} \right).
\end{equation}

\subsection{Exactly solvable model; noninteracting case}

In this subsection, in the multiple free fields case,
we give the Bardeen parameter and the entropic perturbations
and the gravitational wave perturbations and their spectral indices.
Since we can obtain the analytic expression of $\phi_a (\tau)$,
all the results obtained in this subsection have no black boxes
originating from the S formulae connecting the amplitudes of
the final physical quantities and those of the initial physical 
quantities, which were unknown in the paper
\cite{Byrnes2006}.

We consider the exactly solvable model defined by
\begin{equation}
 U = \sum_a \frac{1}{2} m^2_a \phi^2_a,
\label{freescalarfield}
\end{equation}
and calculate the various physical quantities.
The $N$ potential is calculated as
\begin{equation}
 \bar{N} = \frac{\kappa^2}{4} \sum_a 
 \phi^2_a (0),
\end{equation}
then the Bardeen parameter $\zeta_n$ at the end of 
the slow rolling phase is calculated by (\ref{Bardeenleading}).
From now on, we consider the linear perturbations, 
and calculate their spectral indices.
We consider the entropy perturbation between $\phi_a$ and
$\phi_b$ defined by
\begin{equation}
 S_{ab} := - 3 H \left( \frac{1}{\dot{\rho}_a } D \rho_a 
- \frac{1}{\dot{\rho}_b } D \rho_b
\right).
\end{equation}
In the present case (\ref{freescalarfield}), $S_{ab}$ is 
given by
\begin{equation}
 S_{ab} = 3 \kappa^2 U \left(
 \frac{1}{m^2_a \phi_a (0)} 
 \frac{\partial \phi_a (0)}{\partial \lambda}
-\frac{1}{m^2_b \phi_b (0)} 
 \frac{\partial \phi_b (0)}{\partial \lambda}
\right).
\end{equation}
In order to calculate spectral indices, we have to know 
the amplitudes of the quantum fluctuations of the scalar fields
and the gravitational waves at the first 
horizon crossing $\tau = 0$:
\begin{eqnarray}
 \left<\left< \frac{\partial \phi_a (0)}{\partial \lambda}
    \frac{\partial \phi_b (0)}{\partial \lambda}  
 \right>\right>
&\sim& H^2 \delta_{ab} \sim \kappa^2 U \delta_{ab},\\
 \frac{1}{\kappa^2}
 \left<\left< 
    \frac{\partial \tilde{\gamma}_{ij} (0)}{\partial \lambda}
    \frac{\partial \tilde{\gamma}_{ij} (0)}{\partial \lambda}  
 \right>\right>
&\sim& H^2 \sim \kappa^2 U. 
\end{eqnarray}
When we calculate the above correlation functions,
for simplicity we do not take into account the first order
slow rolling corrections unlike the paper \cite{Byrnes2006},
so the above correlation functions become diagonal.
From the horizon crossing relation as
\begin{equation}
 k = a H = \frac{\kappa}{\sqrt{3}} e^N U^{1/2},
\end{equation}
we obtain
\begin{eqnarray}
 d \ln k &=& \left( 
 \kappa^2 U + \frac{1}{2 U} \frac{d U}{d \tau} \right)
 d \tau \notag\\
 &=& \kappa^2 U \left\{ 1 + O \left( \frac{m^2}{H^2_0} \right) 
                \right\} d \tau \notag\\
 &\approx& \kappa^2 U d \tau.
\end{eqnarray}
Then we can calculate the spectral indices as 
\begin{equation}
 \frac{\partial}{\partial \ln{k}} \ln << \zeta^2_1 >>
 = - \frac{4}{\kappa^2} \frac{1}{B(2)}
 \left( \frac{B(2)}{B(0)} + \frac{B(4)}{B(2)} \right),
\end{equation}
\begin{equation}
 \frac{\partial}{\partial \ln{k}} \ln << S^2_{a b} >>
 = \frac{4}{\kappa^2} \frac{1}{B(2)}
 \left( m^2_a m^2_b 
 \frac{m^2_a \phi^2_a (0)+  m^2_b \phi^2_b (0)}
 {m^4_a \phi^2_a (0)+  m^4_b \phi^2_b (0)}
 - \frac{B(4)}{B(2)} 
 \right),
\end{equation}
\begin{equation}
 \frac{\partial}{\partial \ln{k}} 
 \ln << \tilde{\gamma}^2_{i j} >>
 = - \frac{4}{\kappa^2} \frac{B(4)}{B(2)^2},
\end{equation}
where
\begin{equation}
 B(2n) := \sum_a \left( m^2_a \right)^n \phi^2_a (0).
\end{equation}
As for the correlation between the adiabatic and 
the entropic perturbations defined by
\begin{equation}
 {\cal C}_{a b} := \frac{ << \zeta_1 S_{ab}>>}
{ \sqrt{<< \zeta^2_1 >>} \sqrt{<< S^2_{ab}>>}  }, 
\end{equation}
we obtain
\begin{equation}
 \frac{\partial}{\partial \ln{k}} \ln {\cal C}_{a b} 
 = \frac{2}{\kappa^2} \frac{1}{B(0)}
- \frac{2}{\kappa^2} \frac{1}{B(2)}
  m^2_a m^2_b 
 \frac{m^2_a \phi^2_a (0)+  m^2_b \phi^2_b (0)}
 {m^4_a \phi^2_a (0)+  m^4_b \phi^2_b (0)}.
\end{equation}
Since we succeed in solving the scalar fields evolutions completely,
our expressions of the spectral indices of the perturbation variables
are expressed in terms of the initial fields values and the masses
of the scalar fields only, unlike the paper \cite{Byrnes2006} where
the author did not solve the scalar fields evolutions and their 
expressions of the spectral indices contain unknown factors   
originating from the scalar fields evolutions.

\subsection{Effect of the nonresonant interactions}

In the previous subsection, we considered the free scalar
fields.
In this subsection, we will consider the interacting case.
All the interacting systems are classified into the nonresonant
cases and the resonant cases.
By the word ``resonance'', we mean all the factors inducing 
the small denominators in the perturbative expansion 
\cite{Arnold1963}.
Therefore the resonance phenomena are not confined 
in the oscillatory dynamical system. 
In fact, the resonance can also occur in the present slow rolling system.
In the nonresonant case, the masses of the scalar fields do not 
satisfy any resonant relations as shown in Proposition $7$
presented in the below.
In the resonant case, the masses of the scalar fields satisfy 
more than one resonant relation.
In the nonresonant case, the Poincar\'e theorem greatly 
simplifies the treatment of the interactions.
In the nonresonant case, all the interaction terms can be removed 
from the evolution equations by performing the suitable transformation
of the field variables. 
This process is called the linearization of the evolution equations.
Since the linearized, that is transformed evolution equations are the 
free fields equations, they can be solved easily.
The effects of all the interaction terms are contained in the 
transformation law of the field variables.
The fact that the masses of the scalar fields do not satisfy
any resonant relations guarantees the convergence of the transformation
law of the field variables. 
In this way, the evolutions of the interacting scalar fields $\phi_a$
can be solved completely in the form of the analytic functions of 
the $\tau$ function.
As for the concrete nonresonant model, we determine the evolutions
of the scalar fields $\phi_a (\tau)$ and the $N$ potential.
By using this $N$ potential, we can calculate the nonlinearity 
parameters $f_{NL}$, $g_{NL}$ representing the non-Gaussianity
of the Bardeen parameter.
These nonlinearity parameters $f_{NL}$, $g_{NL}$ are shown to be
suppressed by the slow rolling parameter.
Therefore it is difficult to observe such small non-Gaussianity.

First we present the proposition about the nonresonant
condition.
In order to express the nonresonant condition,
we introduce several notations.
By $\alpha$, $k$, we represent
\begin{eqnarray}
 \alpha &:=& 
\left( m^2_1, m^2_2, \cdot \cdot \cdot, m^2_{N_S} \right),\\ 
 k &:=&
\left( k_1, k_2, \cdot \cdot \cdot, k_{N_S} \right),
\end{eqnarray}
where all $k_a$ belong to the nonnegative integer set
$\mathbb{Z}_{0+}$:
\begin{eqnarray}
 k_a &\in& \mathbb{Z}_{0+},\\
\mathbb{Z}_{0+} &:=& \{ k \in \mathbb{Z} | k \ge 0 \}.
\end{eqnarray}
By $|k|$, we represent
\begin{equation}
 |k| := \sum_a |k_a|.
\end{equation}
Then the following proposition holds.

\paragraph{Proposition $7$}

\textit{Suppose that for an arbitrary $(a; k) \in 
\{1,2, \cdot \cdot \cdot, N_S \}
\times \mathbb{Z}^N_{0+} $ satisfying $|k| \ge 2$, }
\begin{equation}
 (k \cdot \alpha) - \alpha_a \neq 0
\end{equation}
\textit{ holds and for an arbitrary 
$a \in \{1,2, \cdot \cdot \cdot, N_S \}$, }
\begin{equation}
 m^2_a > 0
\end{equation}
\textit{holds.
Then there exists a positive constant $\delta_m$ satisfying }
\begin{equation}
 | (k \cdot \alpha) - \alpha_a | \ge \delta_m,
\label{nonresonantpositive}
\end{equation}
\textit{ for an arbitrary 
$(a; k) \in \{1,2, \cdot \cdot \cdot, N_S \}
\times \mathbb{Z}^N_{0+} $ satisfying $|k| \ge 2$. }

\textit{For the proof, see Appendix $I$. }

Under the condition of Proposition $7$, we can prove the 
following proposition.

\paragraph{Proposition $8$}

\textit{ We consider the evolution equation }
\begin{equation}
 \frac{\partial}{\partial \tau} \phi_a =
 - \alpha_a \phi_a + \tilde{f}_a (\phi),
\label{originalevolutionalpha}
\end{equation}
\textit{ where $\tilde{f}_a $ is the sum of $m$-th order monomials
of $\phi_a$ ($m \ge 2$) satisfying }
\begin{equation}
 |\tilde{f}_a (\phi)| < \mu_c \frac{\alpha_M}{\phi_0} |\phi|^2,
\end{equation}
\textit{ where $\alpha_M$ is the maximum of $\alpha_a$, 
for $|\phi| \le \phi_0$. 
Under the condition that $\alpha$ satisfies 
(\ref{nonresonantpositive}), there exists a positive constant $R$
such that, for $\varphi$ satisfying $ |\varphi| \le R$, 
there exists the transformation}
\begin{equation}
 \phi_a = \varphi_a + w_a (\varphi),
\label{transformationphi}
\end{equation}
\textit{ where $w_a$ is the sum of $m$-th order monomials of 
$\varphi_a$ ($m \ge 2$) and satisfies }
\begin{equation}
 |w_a (\varphi)| \le \frac{\mu_c}{\phi_0} |\varphi|^2.
\end{equation}
\textit{ By this transformation law(\ref{transformationphi}), 
(\ref{originalevolutionalpha}) is transformed into the linear
differential equation as}
\begin{equation}
 \frac{\partial}{\partial \tau} \varphi_a = -
 \alpha_a \varphi_a.
\label{exactlylinearevo}
\end{equation}

Proposition $8$ is well known as the Poincar\'e theorem.
\cite{Arnold1963}
The proof is given in Appendix $J$.
This theorem can be applied only to the differential equations
containing the linear terms plus the small perturbations.
Since they are not even the Hamiltonian dynamical systems in general,
Proposition $8$ does not cover the anharmonic oscillator. 
It is well known that the evolution equation of the nonlinear 
anharmonic oscillator cannot be transformed into the linear evolution
equation.
But for the present purpose of treating the multiple slow rolling 
scalar fields systems, Proposition $8$ is enough. 
All the terms in $w_a (\varphi)$ in (\ref{transformationphi})
have factors as
\begin{equation}
 \frac{1}{ (k \cdot \alpha) -\alpha_a }. 
\end{equation}
In order to prove the convergence of this transformation
law (\ref{transformationphi}) by the majorant method, 
the nonresonant inequalities (\ref{nonresonantpositive}) are essential.
According to Proposition $8$, by (\ref{transformationphi})
and 
\begin{equation}
 \varphi_a = \varphi_a (0) \exp{[- \alpha_a \tau]},
\end{equation}
the evolutions of the scalar fields 
$\phi_a$ can be expressed as an analytic function of 
$\tau$.

Next we consider the concrete example whose potential $U$ 
is given by
\begin{equation}
 U = \frac{1}{2} m^2_1 \phi^2_1 +
     \frac{1}{2} m^2_2 \phi^2_2 +
     \frac{g}{4} \phi^2_1 \phi^2_2, 
\end{equation}
where $m^2_1$, $m^2_2$ are assumed to be nonresonant 
as in Proposition $7$.
We obtain 
\begin{eqnarray}
 \phi_1 &=& \phi_1 (0) \exp{(- m^2_1 \tau)}
 - \frac{g}{4 m^2_2} \phi_1 (0) \phi^2_2 (0) 
 \left[ \exp{(- m^2_1 \tau)} 
 - \exp{\{- (m^2_1 + 2 m^2_2) \tau)\} }  \right]
\notag\\
&+& \cdot \cdot \cdot,\\
 \phi_2 &=& \phi_2 (0) \exp{(- m^2_2 \tau)}
 - \frac{g}{4 m^2_1} \phi_2 (0) \phi^2_1 (0) 
 \left[ \exp{(- m^2_2 \tau)} 
 - \exp{\{- (m^2_2 + 2 m^2_1) \tau)\} }  \right]
\notag\\
&+& \cdot \cdot \cdot.
\end{eqnarray}
Then we obtain the $N$ potential as
\begin{equation}
 \frac{1}{\kappa^2} \bar{N} = \frac{1}{4} \phi^2_1(0) +
 \frac{1}{4} \phi^2_2(0)
 - \frac{g}{8 (m^2_1 + m^2_2)}
   \phi^2_1(0) \phi^2_2(0) + \cdot \cdot \cdot.
\end{equation}
In this model, we consider the nonlinear parameters $f_{NL}$,
$g_{NL}$ \cite{Komatsu2001} defined by
\begin{eqnarray}
 f_{NL} &=& \frac{5}{6} 
 \frac{ \bar{N}_{ab} \bar{N}^a \bar{N}^b }
 { \left(\bar{N}_a \bar{N}^a \right)^2 },\\
 g_{NL} &=& \frac{25}{54}
 \frac{ \bar{N}_{abc} \bar{N}^a \bar{N}^b \bar{N}^c }
 { \left(\bar{N}_a \bar{N}^a \right)^3 },
\end{eqnarray}
where 
\begin{equation}
 \bar{N}_a := \frac{\partial}{\partial \phi_a (0)} \bar{N}  
,\quad \quad  
 \bar{N}_{ab} := 
 \frac{\partial^2}{\partial \phi_a (0) \partial \phi_b (0)} 
 \bar{N}, \cdot \cdot \cdot.
\end{equation}
We obtain
\begin{eqnarray}
 f_{NL} &=& \frac{5}{12} \frac{1}{\bar{N}} 
 \sim \left( \frac{m}{H_0} \right)^2 ,\\
 g_{NL} &=& - \frac{25}{144}
\frac{g \kappa^2}{m^2_1 + m^2_2}
\phi^2_1 (0) \phi^2_2 (0) \frac{1}{\bar{N}^3}
\sim \frac{U_{\rm int}}{U}
\left( \frac{m}{H_0} \right)^4.
\end{eqnarray}
So we can conclude that the nonlinear parameters
$f_{NL}$, $g_{NL}$ are suppressed by the slow rolling
parameter $(m / H_0)^2$. 
For an arbitrary potential model satisfying the assumptions 
(i), (ii) and (iii) presented above Proposition $6$ and 
for the nonresonant masses as explained in Proposition $7$, 
the $N$ potential can be written as
\begin{equation}
 \frac{1}{\kappa^2} \bar{N} = \frac{1}{4} \sum_a \phi^2_a (0)
 + \tilde{g} (\phi(0)),
\end{equation}
where $\tilde{g}$ is the sum of $m$-th order monomials of 
$\phi_a (0)$ ($m \ge 3$) and satisfies
\begin{equation}
 \tilde{g} (\phi(0)) \le \mu_c \frac{1}{\phi_0} 
 |\phi(0)|^3,
\end{equation}
for $|\phi(0)| \le \phi_0$.
Then we obtain the nonlinear parameters as
\begin{equation}
 f_{NL} = \frac{5}{12} \frac{1}{\bar{N}} 
 \sim \left( \frac{m}{H_0} \right)^2 ,\quad \quad
 g_{NL} \sim \mu_c \left( \frac{m}{H_0} \right)^4,
\end{equation}
that is, the nonlinear parameters $f_{NL}$, $g_{NL}$ are 
suppressed by the slow rolling parameters $(m / H_0)^2$.

\subsection{Effect of the resonant interactions}

According to Proposition $8$ in the previous subsection, 
all the nonresonant interaction terms can be eliminated 
by the field transformations $\phi_a \to \varphi_a$.
But in the resonant interaction case, such linearization
cannot be applied.
In this subsection, we calculate the effect of the resonant 
interaction by the iteration method.
Perturbatively at least, the resonant interactions do not 
generate any special effects in the $N$ potential.

As the concrete example, we consider the model defined by 
\begin{equation}
 U_{\rm int} = \lambda \phi_1 \phi^2_2, \quad \quad
 m^2_1 = 2 m^2_2.
\end{equation}
It can be solved as
\begin{eqnarray}
 \phi_1 &=& \phi_1 (0) \exp{(- m^2_1 \tau)}
 - \lambda \phi^2_2 (0) \tau
   \exp{(- m^2_1 \tau)}
 + \cdot \cdot \cdot,\\
 \phi_2 &=& \phi_2 (0) \exp{(- m^2_2 \tau)}
 - \frac{2 \lambda}{m^2_1} \phi_1 (0) \phi_2 (0) 
 \left[ \exp{(- m^2_2 \tau)} 
 - \exp{\{- (m^2_1 + m^2_2) \tau)\} }  \right]
\notag\\
&+& \cdot \cdot \cdot.
\end{eqnarray}
where in the right hand side of $\phi_1$ 
the second term proportional to $\lambda \tau$ appears
because of the resonant interaction.
The $N$ potential is given by
\begin{equation}
 \frac{1}{\kappa^2} \bar{N} = \frac{1}{4} \phi^2_1(0) +
 \frac{1}{4} \phi^2_2(0)
 - \frac{5 \lambda}{8 m^2_2}
   \phi_1(0) \phi^2_2(0) + \cdot \cdot \cdot.
\end{equation}
So the resonant interaction terms generate 
no singular terms in the $N$ potential $\bar{N}$.

\section{Discussion}

We presented the nonlinear LWL formula and by using it we
investigated the evolutionary behaviors of the nonlinear 
cosmological perturbations on superhorizon scales in the 
universe dominated by the multiple slow rolling scalar fields.

In the excellent study as to the multiple slow rolling scalar 
fields \cite{Gordon2001} \cite{Byrnes2006}, 
the multiple scalar fields were decomposed into the adiabatic
field and the entropy fields instant by instant.
By using this decomposition, it was pointed out 
that the growing of the Bardeen parameter corresponds to 
the curvature of the trajectory in the scalar fields space.
Further in the paper \cite{Byrnes2006}, the spectral index 
(the wavenumber dependences) of the Bardeen parameter 
was presented by calculating the differential 
coefficients at the instant of the first horizon crossing.
But this study is the local investigation because the author 
did not calculate the S formulae connecting the final amplitudes 
of the adiabatic and the entropic fields variables at the 
end of the slow rolling phase with the initial amplitudes of them
at the first horizon crossing, and they calculated only 
the time derivative of the S formulae at the first
horizon crossing by using the perturbation evolution equations.
So the formulae of the spectral indices in the paper \cite{Byrnes2006}
have black boxes (unknown factors) which come from the S formulae 
which could not be determined.
Therefore as the method for investigating the 
evolutions of the scalar fields perturbations in the long time 
interval such as in the whole slow rolling phase analytically, 
we present the method using the $\tau$ function and the $N$ potential.
This method enables us to calculate the S formulae analytically. 
In fact, our expressions of spectral indices in subsection $3.3$
can be expressed in terms of the initial field values and the masses
of the scalar fields only and they have no blackboxes  
since we succeed in calculating the scalar fields evolutions 
containing the S formulae completely.
Our method will deepen the understanding of the dynamical evolutions 
of the multiple slow rolling scalar fields in the long time interval
with high accuracy.

\section*{Acknowledgments}

The author thanks Professors H. Kodama, M. Sasaki, J. Yokoyama 
for continuous encouragements.
He thanks the referee for making him notice that $RT$ 
in (\ref{correctionsolution}) is symmetric.

\appendix



\catcode`\@=11

\@addtoreset{equation}{section}   
\def\theequation{\Alph{section}.\arabic{equation}}


\section{Derivation of (\ref{constantconstraint})}

In this appendix, we rewrite the momentum constraints 
(\ref{momentumconstraint}) into the constraints which the 
solution constants of the solution of the locally homogeneous 
universe must satisfy.

By using (\ref{extrinsictracepart})(\ref{extrinsictracelessdef}),
the left hand side of the momentum constraint (\ref{momentumconstraint})
is expressed as
\begin{eqnarray}
 && D_i \tilde{A}^{i}_{\> j} -\frac{2}{3} D_j K 
\notag\\
&=& \frac{1}{\alpha}
\left[     
 \frac{1}{2} \frac{\partial_i \alpha}{\alpha} 
 \tilde{\gamma}^{i k} \partial_t \tilde{\gamma}_{k j}
 + \frac{1}{2} 
 \tilde{\gamma}^{i l} \partial_i \tilde{\gamma}_{l m} 
 \tilde{\gamma}^{m k} \partial_t \tilde{\gamma}_{k j}
 - \frac{1}{2}
 \tilde{\gamma}^{i k} \partial_i \partial_t \tilde{\gamma}_{k j}
\right.
\notag\\
&&
\left. -
 \frac{3}{2} \frac{\partial_k a}{a} 
 \tilde{\gamma}^{k l} \partial_t \tilde{\gamma}_{l j}
+ \frac{1}{2} {^s \tilde{\Gamma}^k_{i j} }
 \tilde{\gamma}^{i l} \partial_t \tilde{\gamma}_{l k}
+ 2 \alpha \partial_j \left( \frac{H}{\alpha} \right)
\right],
\label{momentumleft}
\end{eqnarray}
where ${^s \tilde{\Gamma}^k_{i j} }$ is the Christoffel 
symbol of $\tilde{\gamma}_{i j}$.
As for the right hand side of (\ref{momentumleft}),
\begin{equation}
 \frac{1}{2} \frac{\partial_i \alpha}{\alpha} 
 \tilde{\gamma}^{i k} \partial_t \tilde{\gamma}_{k j}
 - \frac{3}{2} \frac{\partial_k a}{a} 
 \tilde{\gamma}^{k l} \partial_t \tilde{\gamma}_{l j}
= - \frac{1}{2} \frac{\alpha}{a^3}
 \partial_i \left( \frac{a^3}{\alpha} \right)
 \left( M^{-1} \dot{M} \right)^i_{\> j},
\end{equation}
and
\begin{equation}
 {^s \tilde{\Gamma}^k_{i j} }
 \tilde{\gamma}^{i l} \partial_t \tilde{\gamma}_{l k}
 = \frac{1}{2} {\rm tr} 
 \left[ M^{-1} \partial_j M M^{-1} \dot{M} \right],
\end{equation}
and
\begin{equation}
 \tilde{\gamma}^{i l} \partial_i \tilde{\gamma}_{l m} 
 \tilde{\gamma}^{m k} \partial_t \tilde{\gamma}_{k j}
 = \left(   
 M^{-1} \partial_i M M^{-1} \dot{M}
   \right)^i_{\> j},
\end{equation}
and
\begin{equation}
  \tilde{\gamma}^{i k} \partial_i \partial_t \tilde{\gamma}_{k j}
 = \left( M^{-1} \partial_i \dot{M}  \right)^i_{\> j}.
\end{equation}
Therefore we obtain
\begin{eqnarray}
 && D_i \tilde{A}^{i}_{\> j} -\frac{2}{3} D_j K \notag\\
&=& \frac{1}{\alpha} 
\left[ - \frac{1}{2} \frac{\alpha}{a^3} \partial_i T^i_{\> j}
  + \frac{1}{4} \frac{\alpha}{a^3}  
  {\rm tr} \left( R^{-1} \partial_j R T \right) 
\right.\notag\\
&& \left. + 
 \frac{\alpha}{a^3} \int_{t_0} dt \partial_j 
 \left( \frac{\alpha}{a^3} \right) c_T
+ \frac{1}{2} \frac{\alpha}{a^3} \int_{t_0} dt
 \frac{\alpha}{a^3} \partial_j c_T
+ 2 \alpha \partial_j \left( \frac{H}{\alpha} \right)
\right],
\end{eqnarray}
where we used the solution of $M$ (\ref{correctionsolution}).

The right hand side of (\ref{momentumconstraint}) 
is expressed as
\begin{equation}
 \kappa^2 J_j = \frac{1}{\alpha} 
 [ \kappa^2 h (v_j + \beta_j)].
\label{momentumcurrent}
\end{equation}
By summing (\ref{integratevelocityalpha}) with fluid indices
$\alpha$, we obtain 
\begin{equation}
  \left[ h (\beta_i + v_i) \right]_f =
 \frac{\alpha}{a^3} \sum_{\alpha} C_{\alpha i} +
 \frac{\alpha}{a^3} \int_{t_0} dt \alpha a^3
 \left[
- \partial_i P_f - \frac{1}{\alpha} D_i \alpha h_f 
- \sum_a S_a \partial_i \phi_a
 \right]  + O(\epsilon^3),
\label{velocityfluidsum}
\end{equation}
where we used $\sum_{\alpha} Q_{\alpha i} = - (Q_i)_S = 
- \sum_a S_a \partial_i \phi_a$
and as for the scalar fields components we
obtain
\begin{equation}
 \left[ h_a (\beta_i + v_{a i}) \right]_S 
 = - \sum_a \dot{\phi}_a \partial_i \phi_a
 + O(\epsilon^3).
\label{velocityscalarsum}
\end{equation}
We substitute the sum of (\ref{velocityfluidsum}) and 
(\ref{velocityscalarsum}) into $h (\beta_i + v_i)$ in 
(\ref{momentumcurrent}).

Then through simple calculations we obtain 
(\ref{constantconstraint}).

\section{Lie derivative in the leading order of 
the gradient expansion }

The Lie derivatives of the quantity of upper index
and the quantity with lower index are expressed as
\begin{eqnarray}
 L(T) X^{\mu} &=& T^{\alpha} \partial_{\alpha} X^{\mu} 
 - \partial_{\alpha} T^{\mu} X^{\alpha},\\
 L(T) X_{\mu} &=& T^{\alpha} \partial_{\alpha} X_{\mu}
 + X_{\alpha} \partial_{\mu} T^{\alpha},
\end{eqnarray}
respectively.
By the Leibniz rule, these definitions are expanded into the tensor 
of an arbitrary rank.
For example, the Lie derivative of the metric is given by
\begin{equation}
  L(T) g_{\mu \nu} = T^{\alpha} \partial_{\alpha} g_{\mu \nu}
 + g_{\alpha \nu} \partial_{\mu} T^{\alpha}
 + g_{\mu \alpha} \partial_{\nu} T^{\alpha}.
\label{Liemetric}
\end{equation}
From now on, we consider the gradient expansion scheme
defined by
\begin{equation}
 \partial_i = O(\epsilon), \quad \quad
 T^i = O(\epsilon), \quad \quad
 g_{i 0} = O(\epsilon), 
\label{gradorder}
\end{equation}
and the $O(\epsilon^2)$ order corrections are dropped.
For an arbitrary scalar quantity $A$, in our scheme (\ref{gradorder})
the Lie derivative of $A$ is written by
\begin{equation}
 L(T) A = T^0 \frac{d}{d t} A + O(\epsilon^2).
\label{Liescalar}
\end{equation}
By using (\ref{Liemetric}) to $g_{00} = - \alpha^2 + \beta_k \beta^k$
and $g_{i j} = \gamma_{i j}$, we obtain
\begin{eqnarray}
 L(T) \alpha &=& T^0 \frac{d}{d t} \alpha 
 + \alpha \frac{d}{d t} T^0 + O(\epsilon^2),
\label{Lielapse}\\
 L(T) \gamma_{i j} &=& T^0 \frac{d}{d t} \gamma_{i j} 
+ O(\epsilon^2).
\label{Liespacemetric}
\end{eqnarray}
From (\ref{Liescalar}) and (\ref{Lielapse}), we obtain
\begin{equation}
 L(T) \left( \frac{\dot{A}}{\alpha} \right) = 
 T^0 \frac{d}{d t} \left( \frac{\dot{A}}{\alpha} \right)+ O(\epsilon^2).
\label{Liescalartimeder}
\end{equation}
From this equation, we can see that the Lie derivative of
$\dot{A} / \alpha$ where $A$ is a scalar quantity has 
the same form as the Lie derivative of the scalar quantity
(\ref{Liescalar}).
By using (\ref{Liespacemetric}) to ${\rm det} (\gamma_{i j}) = a^6$,
we obtain 
\begin{equation}
 L(T) a = T^0 \frac{d}{d t} a + O(\epsilon^2),
\end{equation}
therefore we obtain
\begin{equation}
  L(T) \tilde{\gamma}_{i j} = T^0 \frac{d}{d t} \tilde{\gamma}_{i j} 
+ O(\epsilon^2).
\end{equation}
In our gradient expansion scheme, the scale factor $a$ and
$\tilde{\gamma}_{i j}$ can be regarded as the scalar quantity.

\section{Proof of Proposition $1$ }

The gauge transformation is described by the differential 
equation (\ref{gaugetransdiffeq}).
So we can consider the $\mu$ derivative as the gauge transformation.
By differentiating (\ref{gaugetransdiffeq}) with respect to $\lambda$,
we obtain
\begin{eqnarray}
 \frac{d}{d \mu} \frac{d A}{d \lambda} &=&
 L \left( \frac{d T}{d \lambda} \right) A +
 L(T) \frac{d A}{d \lambda},\\
 \frac{d}{d \mu} \frac{d^2 A}{d \lambda^2} &=&
 L \left( \frac{d^2 T}{d \lambda^2} \right) A +
 2 L \left( \frac{d T}{d \lambda} \right) \frac{d A}{d \lambda}
 + L(T) \frac{d^2 A}{d \lambda^2},
\end{eqnarray}
and so on.
As for $D^n A$ where $D$ is defined by (\ref{Doperation}) and 
$A$ is an arbitrary scalar quantity, we can prove
\begin{equation}
 \frac{d}{d \mu} D^n A = L(T) D^n A,
\label{muderDNA}
\end{equation}
for an arbitrary natural number $n$.

\textit{Proof}

If we interpret $D^0 A := A$, (\ref{muderDNA}) holds evidently 
for $n=0$.
We assume that (\ref{muderDNA}) holds for $n = k-1$ where
$k = 1, 2, \cdot \cdot \cdot$.
Then for $n=k$
\begin{eqnarray}
 \frac{d}{d \mu} D^k A &=& \frac{d}{d \mu} \left\{
\left( \frac{d}{d \lambda} - 
 \frac{d a}{d \lambda} \frac{1}{\dot{a}} \frac{d}{d t}
              \right) D^{k-1} A \right\}
\notag\\
&=& \frac{d}{d \lambda} \frac{d}{d \mu} D^{k-1} A
- \frac{d}{d \lambda} \left( \frac{d a}{d \mu} \right)
 \frac{1}{\dot{a}}\left( D^{k-1} A \right)^{\cdot}
- \frac{d a}{d \lambda} \frac{d}{d \mu}
\left\{     
  \frac{1}{\dot{a}}\left( D^{k-1} A \right)^{\cdot}
\right\}
\notag\\
&=&
L \left( \frac{d T}{d \lambda} \right) D^{k-1} A
+ L (T) \frac{d}{d \lambda} D^{k-1} A
- \left\{    
  L \left( \frac{d T}{d \lambda} \right) a
+ L (T) \frac{d a }{d \lambda} 
  \right\}
 \frac{1}{\dot{a}}\left( D^{k-1} A \right)^{\cdot}
\notag\\
&&
- \frac{d a}{d \lambda} L(T)
 \left\{  \frac{1}{\dot{a}}\left( D^{k-1} A \right)^{\cdot} 
 \right\}
\notag\\
&=& L(T) \left\{
\left( \frac{d}{d \lambda} - 
 \frac{d a}{d \lambda} \frac{1}{\dot{a}} \frac{d}{d t}
              \right) D^{k-1} A \right\}
\notag\\
&=& L(T) D^k A
\end{eqnarray}
where we use the fact that both $D^{k-1} A$ and the scale factor $a$
have the Lie derivatives of the scalar quantity type
(\ref{Liescalar}).
For $n=k$, (\ref{muderDNA}) holds.
By induction, we complete the proof. $\blacksquare$

By putting $\lambda = 0$ in (\ref{muderDNA}) and by noticing
$T(\lambda = 0)=0$
\begin{equation}
 \frac{d}{d \mu} D^n A \Bigg|_{\lambda=0} = 0.
\end{equation}
Then we proved that $D^n A$ where $A$ is an arbitrary scalar
quantity is gauge invariant.
Since $\dot{A} / \alpha$ where $A$ is an arbitrary scalar
quantity has the same Lie derivative 
as that of the scalar quantity $A$, 
$D^n (\dot{A} / \alpha)$ is also gauge invariant.
In the way similar to the above calculation, as for ${\cal A}_n$
defined by (\ref{calAN}) we can prove
\begin{equation}
 \frac{d}{d \mu} {\cal A}_n = L(T) {\cal A}_n.
\end{equation}
Then ${\cal A}_n$ is gauge invariant.

\section{Proof of Proposition $2$ }

Since $\zeta_1$ is conserved for arbitrary values of the 
integration constants $C(\lambda = 0)$,  
$\zeta_1$ with $C(\lambda = 0)$ replaced with $C(\lambda)$
is also conserved, that is, $\zeta_1$ is conserved for arbitrary 
values of $\lambda$:
\begin{equation}
 \left( \frac{\partial}{\partial a} \zeta_1 \right)_a = 0.
\end{equation}
Then by the expression of $\bar{\delta}$ 
(\ref{deltabarscalefactor}) we obtain
\begin{equation}
 \zeta_2 = \left(
 \frac{\partial}{\partial \lambda} \zeta_1 \right)_a.
\end{equation}
Since $(\partial / \partial a)_a$ and $(\partial / \partial \lambda)_a$
are commutative, we obtain
\begin{equation}
 \left( \frac{\partial}{\partial a} \zeta_2 \right)_a =
 \left( \frac{\partial}{\partial a}
  \frac{\partial}{\partial \lambda} \zeta_1 \right)_a =
 \left( \frac{\partial}{\partial \lambda} 
 \frac{\partial}{\partial a} \zeta_1 \right)_a =0.
\end{equation}
By iterating the same process, we obtain
\begin{equation}
 \zeta_n = \left( \frac{\partial}{\partial \lambda} \right)^{n-1}_a
 \zeta_1, \quad \quad 
 \left( \frac{\partial}{\partial a} \zeta_n \right)_a =0,
\end{equation}
for $n \ge 2$.

\section{Proof of Proposition $3$ }

When $P = P(\rho)$,
\begin{equation}
 a \frac{\partial \rho}{\partial a} = - 3 (\rho + P). 
\end{equation}
Then 
\begin{eqnarray}
 \zeta_1 &=& - 
 \left\{ \left(\frac{\partial \rho}{\partial \lambda} \right)_a \Big/ 
         \left(\frac{\partial \rho}{\partial a} \right)_a \right\}
 \frac{1}{a} \notag\\
 &=& \frac{1}{3} \frac{1}{\rho + P} 
\left( \frac{\partial \rho}{\partial \lambda} \right)_a \notag\\
&=& \left( \frac{\partial}{\partial \lambda} \right)_a
  \left[ \frac{1}{3} 
 \int d \rho \frac{1}{\rho + P(\rho)} 
  \right],
\end{eqnarray}
and 
\begin{eqnarray}
 \left( \frac{\partial}{\partial a} \zeta_1 \right)_a
 &=& \left( \frac{\partial}{\partial \lambda}
     \frac{\partial}{\partial a} \right)_a
\left[ \frac{1}{3} \int d \rho \frac{1}{\rho + P(\rho)} \right]
\notag\\
&=& \left( \frac{\partial}{\partial \lambda} \right)_a
 \left( - \frac{1}{a} \right)
\notag\\
&=& 0.
\end{eqnarray}
since $(\partial / \partial \lambda)_a$ and $(\partial / \partial a)_a$
are commutative.
Then we can use Proposition $2$, we complete the proof.

\section{Proof of Proposition $4$ }

For $k=0$, by solving (\ref{evolutionnewphi}) (\ref{evolutionnewp})
we obtain
\begin{eqnarray}
 |p - e^{-3N} p(0)| &\le& \eta^2,\\
 |\phi - \phi(0)| &\le& \eta^2 N.
\end{eqnarray}
If $p(0), \phi(0) \sim 1$, Proposition $4$ is right for $k=0$.
We assume that Proposition $4$ holds for 
$k=0,1,2, \cdot \cdot \cdot ,k-1$:
\begin{equation}  
 \left| \left( \frac{\partial^i}{\partial \lambda^i} \right)_a 
 \phi \right|, \quad
 \left| \left( \frac{\partial^i}{\partial \lambda^i} \right)_a 
 p \right|
 \sim \delta^i_c, \quad \quad (i=0,1,2, \cdot \cdot \cdot, k-1),
\end{equation}
therefore
\begin{equation}
 \left| \left( \frac{\partial^i}{\partial \lambda^i} \right)_a 
 f(\phi, p, c_T, N) \right|
 \sim \delta^i_c, \quad \quad (i=0,1,2, \cdot \cdot \cdot, k-1),
\end{equation}
if 
\begin{equation}
 | f(\phi, p, c_T, N) | \le 1, 
\end{equation}
for a proper complex domain containing the real interval which we
consider.
For $k$, the evolution equations are
\begin{eqnarray}  
 \frac{d}{d N} \left( 
 \frac{\partial^k}{\partial \lambda^k} \right)_a \phi 
&=& \eta^2 \left( 
 \frac{\partial^k}{\partial \lambda^k} \right)_a \phi
  + \eta^2 \left( 
 \frac{\partial^k}{\partial \lambda^k} \right)_a p
  + \eta^2 \delta^k_c,
\label{kphi}\\
  \frac{d}{d N} \left( 
 \frac{\partial^k}{\partial \lambda^k} \right)_a p
&=& - 3 \left( 
  \frac{\partial^k}{\partial \lambda^k} \right)_a p
  + \eta^2 \left( 
  \frac{\partial^k}{\partial \lambda^k} \right)_a \phi
  + \eta^2 \left( 
  \frac{\partial^k}{\partial \lambda^k} \right)_a p
  + \eta^2 \delta^k_c,
\label{kp}
\end{eqnarray}
where all the coefficients bounded by positive constants are 
omitted except $-3$ in (\ref{kp}).
By solving (\ref{kphi})(\ref{kp}), for $0 \le N \le 1/ \eta^2$,
we obtain
\begin{eqnarray} 
 \left| \left( 
  \frac{\partial^k}{\partial \lambda^k}  \right)_a \phi \right|
&\le& \left| \frac{\partial^k}{\partial \lambda^k} \phi (0) \right|
 + \eta^2 \left| \frac{\partial^k}{\partial \lambda^k} p(0) \right|
 + \delta^k_c,\\
 \left| \left( 
 \frac{\partial^k}{\partial \lambda^k}  \right)_a p \right|
&\le& e^{- 3 N} \left| \frac{\partial^k}{\partial \lambda^k} p(0) \right|
 + \eta^2 \left[ 
 \left| \frac{\partial^k}{\partial \lambda^k} \phi (0) \right|
 + \eta^2 \left| \frac{\partial^k}{\partial \lambda^k} p(0) \right|
 + \delta^k_c 
\right],
\end{eqnarray}
Therefore for the initial conditions (\ref{initialpartialk}),
for $k$, (\ref{partialkphi})(\ref{partialkp}) hold.
By induction, for all nonnegative integers $k$, Proposition $4$
holds.

\section{Proof of Proposition $5$ }

First we consider $k = 0$ case.
By the mean value theorem, we obtain
\begin{equation}
 \frac{d}{d N} \Delta \phi = \eta^2 \Delta \phi
 + \eta^2 f,
\label{errorofphi}
\end{equation}
where 
\begin{equation}
 |f| \le |p| + \eta^2 \le e^{- 3 N} + \eta^2,
\end{equation}
where we used Proposition $4$ in the last inequality.
By solving (\ref{errorofphi}), we obtain
\begin{equation}
 |\Delta \phi| \le \exp{(\eta^2 N)} 
 \left( |\Delta \phi(0)| + \eta^2 + \eta^4 N \right).
\end{equation}
Therefore for $0 \le N \le 1 / \eta^2$, under the initial condition
(\ref{initialerror}), (\ref{partialkerror}) holds for $k=0$.
Next we consider a positive integer $k$ case.
In this case, we notice the following fact.
By Proposition $4$, if an complex analytic function $f$ satisfies
\begin{equation}
 |f(\phi, p, c_T, N)| \le 1,
\end{equation}
then 
\begin{equation} 
 \left| \left( \frac{\partial^k}{\partial \lambda^k} \right)_a
 f(\phi, p, c_T, N) \right| \le \delta^k_c,
\end{equation}
By differentiating (\ref{errorofphi}) with respect to $\lambda$
and by using the above fact and Proposition $4$, we obtain
\begin{equation}
 \frac{d}{d N} \left( 
 \frac{\partial^k}{\partial \lambda^k} \right)_a \Delta \phi
 = \eta^2 \left( 
 \frac{\partial^k}{\partial \lambda^k} \right)_a \Delta \phi
 + \eta^2 \delta^k_c \left( e^{- 3 N} + \eta^2 \right),
\end{equation}
which is solved as 
\begin{equation}
 \left| \left( 
  \frac{\partial^k}{\partial \lambda^k} \right)_a \Delta \phi \right| 
\le \exp{(\eta^2 N)} 
 \left( \left| \frac{\partial^k}{\partial \lambda^k} \Delta \phi(0) \right| 
 + \eta^2 \delta^k_c + \eta^4 N \delta^k_c \right).
\end{equation}
Therefore for $0 \le N \le 1 / \eta^2$, under the initial condition
(\ref{initialerror}), (\ref{partialkerror}) holds for a positive 
integer $k$.

\section{Proof of Proposition $6$ }

\paragraph{Lemma $1$}
\begin{equation}
 \phi_a = \phi_a (0) \exp{[- m^2_a \tau]}
 + < \mu_c \phi_0 \exp{[- m^2 \tau]}>.
\end{equation}

\paragraph{Proof}
We consider the evolution equation
\begin{equation}
 \frac{\partial}{\partial \tau} \phi_a =
 - m^2_a \phi_a  + \tilde{f}_a (\phi),
\label{evolutionphitau}
\end{equation}
where $\tilde{f}_a (\phi)$ is the sum of $m$-th order 
monomials ($m \ge 2$) satisfying
\begin{equation}
 | \tilde{f}_a (\phi) | \le m^2 \mu_c \frac{1}{\phi_0}
 |\phi|^2,
\end{equation}
for $|\phi| \le \phi_0$.
From (\ref{evolutionphitau}), we obtain 
\begin{equation}
 \frac{\partial}{\partial \tau} |\phi| \le
 - m^2 |\phi|  + \frac{m^2 \mu_c}{\phi_0} |\phi|^2,
\end{equation}
which is solved as 
\begin{equation}
 |\phi| \le \phi_0 \exp{[- m^2 \tau]} (1+O(\mu_c)).
\label{absphiestimate}
\end{equation}
We consider $\bar{\phi}_a$ satisfying
\begin{eqnarray}
  \frac{\partial}{\partial \tau} \bar{\phi}_a &=&
 - m^2_a \bar{\phi}_a,\\
 \bar{\phi}_a (0) &=& \phi_a (0).
\end{eqnarray}
The difference $\Delta \phi_a := \phi_a - \bar{\phi}_a$ 
satisfies
\begin{eqnarray}
  \frac{\partial}{\partial \tau} \Delta \phi_a &=&
 - m^2_a \Delta \phi_a + \tilde{f}_a (\phi),\\
 \Delta \phi_a (0) &=& 0,
\end{eqnarray}
where 
\begin{equation}
 |\tilde{f}_a (\phi)| \le m^2 \mu_c \phi_0 \exp{[- m^2 \tau]},
\end{equation}
where we used (\ref{absphiestimate}).
By solving the above evolution equation under the above 
initial condition, we obtain
\begin{equation}
 |\Delta \phi| \le O(\mu_c) \phi_0 \exp{[- m^2 \tau]}.
\end{equation}
We complete the proof.
$\blacksquare$

\paragraph{Lemma $2$}
\begin{equation}   
 \left( \frac{\partial^k}{\partial \lambda^k} \right)_{\tau} \phi_a =
 \frac{\partial^k}{\partial \lambda^k} \phi_a (0) \cdot
 \exp{[- m^2_a \tau]} +
 <\delta^k_c \mu_c \phi_0 \exp{[- m^2 \tau]}>.
\end{equation}

\paragraph{Proof}
As for the evolution equation as 
\begin{equation}
 \frac{\partial}{\partial \tau} 
 \left( \frac{\partial^k}{\partial \lambda^k} \right)_{\tau} \phi_a =
 - m^2_a \left( 
 \frac{\partial^k}{\partial \lambda^k} \right)_{\tau} \phi_a  
 + \left( \frac{\partial^k}{\partial \lambda^k} \right)_{\tau} 
   \tilde{f}_a (\phi),
\end{equation}
where 
\begin{eqnarray}
 \left| \left( \frac{\partial^k}{\partial \lambda^k} \right)_{\tau} 
 \tilde{f}_a (\phi)
 \right| &\le&
 m^2 \frac{\mu_c}{\phi_0} |\phi|
 \left| \left( \frac{\partial^k}{\partial \lambda^k} \right)_{\tau}
 \phi_a 
 \right|
 +  m^2 \frac{\mu_c}{\phi_0} \delta^k_c |\phi|^2
\notag\\
&\le& m^2 \mu_c \exp{[- m^2 \tau]}
  \left| \left( \frac{\partial^k}{\partial \lambda^k} \right)_{\tau}
  \phi_a 
 \right|
 + \delta^k_c \mu_c m^2 \phi_0 \exp{[- 2 m^2 \tau]}.
\end{eqnarray}
We perform the calculations similar to the proof of
Lemma $1$.
We complete the proof.
$\blacksquare$

\paragraph{Lemma $3$}
\begin{equation}  
 \left( \frac{\partial^k}{\partial \lambda^k} \right)_{\tau}
 \left( \prod^N_{l=1} \phi_{a(l)} \right)
= 
\frac{\partial^k}{\partial \lambda^k} 
 \left( \prod^N_{l=1} \phi_{a(l)} (0) \right)
 \exp{[ - \sum^N_{l=1} m^2_{a(l)} \tau]}
 + <\delta^k_c \mu_c \phi^N_0 \exp{[- N m^2 \tau]}>.
\end{equation}

\paragraph{Proof}

We use the Leibniz rule for the $\lambda$ derivative
and use Lemma $1$, Lemma $2$.
We complete the proof.
$\blacksquare$

\paragraph{Lemma $4$}
\textit{For $n \ge 1$}
\begin{equation}
 A (2 n, 0) = \sum_a \left( m^2_a \right)^n
 \phi^2_a + F_{2 n} (\phi)
\label{polynomialA2n}
\end{equation}
\textit{
where $F_{2 n} (\phi)$ is the sum of $m$-th order 
monomials of $\phi$ ($m \ge 3$) and satisfies }
\begin{equation}
 |F_{2 n} (\phi)| \le \mu_c m^{2 n} \frac{1}{\phi_0}
 |\phi|^3
\end{equation}
\textit{for $|\phi| \le \phi_0$. }

\paragraph{Proof}
We use (\ref{evolutionphitau}).
We complete the proof.
$\blacksquare$

As for 
\begin{equation}  
 \left( \frac{\partial^k}{\partial \lambda^k} \right)_{\tau} 
 \bar{N}
 = \kappa^2 \int^{\infty}_0 d \tau  
 \left( \frac{\partial^k}{\partial \lambda^k} \right)_{\tau} U,
\end{equation}
and 
\begin{equation}
 A(0, k) = 4 \int^{\infty}_{\tau}
 d \tau
 \left( \frac{\partial^k}{\partial \lambda^k} \right)_{\tau} U,
\end{equation}
and 
\begin{equation}
 A(2n, k) =
 \left( \frac{\partial^k}{\partial \lambda^k} \right)_{\tau} 
 A(2n, 0),
\end{equation}
for $n \ge 1$
where $A(2n, 0)$ is expressed as a polynomial of $\phi (\tau)$
(\ref{polynomialA2n}), we use Lemma $3$.
Then we complete the proof of Proposition $6$.

\section{Proof of Proposition $7$}

By $\alpha_M$, $\alpha_m$, we mean
\begin{eqnarray}
 \alpha_M &:=& \max_a \{ m^2_a \},
\label{alphaMdef}\\
 \alpha_m &:=& \min_a \{ m^2_a \}.
\end{eqnarray}
Then we obtain
\begin{equation}
 | (k \cdot \alpha) - \alpha_a | \ge 
 |k| \alpha_m - \alpha_M.
\end{equation}
Therefore for $k$ satisfying
\begin{equation}
 |k| \ge \frac{\delta_1 + \alpha_M}{\alpha_m}
 =: L,
\end{equation}
where $\delta_1$ is a positive constant, we obtain 
\begin{equation}
 | (k \cdot \alpha) - \alpha_a | \ge \delta_1.
\end{equation}
Since the number of $k$ satisfying $|k| < L$ 
is finite, there exists a positive constant $\delta_2$
satisfying
\begin{equation}
 | (k \cdot \alpha) - \alpha_a | \ge \delta_2,
\end{equation}
for $|k| < L$.
Therefore for $\delta_m$ defined by
\begin{equation}
 \delta_m := \min \{ \delta_1, \delta_2 \},
\end{equation}
the inequality (\ref{nonresonantpositive}) holds.
We complete the proof.

\section{Proof of Proposition $8$}

By replacing $\phi_a$ with $\phi_0 \phi_a$ and 
by replacing $\tau$ with $\tau / \mu_c \alpha_M$,
we can assume that $|\tilde{f}_a (\phi)| \le 1$
for $|\phi| \le 1$.
In order that the transformation law (\ref{transformationphi})
can reduce the original equation (\ref{originalevolutionalpha})
to the exactly linear equation (\ref{exactlylinearevo}),
$w_a$ must satisfy
\begin{equation}
 - \frac{\partial w_a}{\partial \varphi_b} \alpha_b
 \varphi_b + \alpha_a w_a 
= \tilde{f}_a (\varphi + w).
\label{wasatisfy}
\end{equation}
We decompose $w_a$ into the sum of the $K$-th order 
polynomials:
\begin{equation}
 w_a = \sum_{K \ge 2} w^K_a, \quad \quad
 w^K_a = \sum_{|k| = K} w^k_a \varphi^k,
\end{equation}
where
\begin{equation}
 \varphi^k := \varphi_1^{k_1} \cdot \cdot \cdot
 \varphi_{N_S}^{k_{N_S}},\quad \quad
 |k| := k_1 + \cdot \cdot \cdot + k_{N_S}.
\end{equation}
We decompose (\ref{wasatisfy}) into $K$-th order 
part as 
\begin{equation}
 - \frac{\partial w^K_a}{\partial \varphi_b} \alpha_b
 \varphi_b + \alpha_a w^K_a 
= P^K_a ( \tilde{f}, w^1, \cdot \cdot \cdot, w^{K-1}),
\quad \quad
 P^K_a = \sum_{|k|=K} d^k_a \varphi^k
\end{equation}
for $K = 2, \cdot \cdot \cdot$.
We mean the coefficient of $\varphi^k$ in $\tilde{f}_a$
by $f^k_a$.
Then $d^k_a$ ($|k|=K$) is the 
polynomial of $f^k_a$ ($|k| \le K$) and 
$w^k_a$ ($|k| \le K-1$) with integral coefficients and 
$w^k_a$ can be expressed as
\begin{equation}
 w^k_a = \frac{d^k_a}{-(k \cdot \alpha)+\alpha_a}.
\end{equation}
From now on, we will prove the convergence of the transformation 
$w_a$ by the method of the majorant.
$\tilde{F} (\varphi)$ defined by
\begin{equation}
 \tilde{F} (\varphi) = \sum_{K=2}^{\infty}
(\varphi_1 + \cdot \cdot \cdot + \varphi_{N_S})^K
\end{equation}
is the majorant of $\tilde{f}_a$.
We consider the equation as
\begin{equation}
 \delta_m \cdot \tilde{W}(\varphi) = 
 \tilde{F}(\varphi+\tilde{W}(\varphi)),
\label{majorantW}
\end{equation}
where $\delta_m$ is defined in Proposition $7$,
Since $\tilde{F} (\varphi)$ is $a$ independent,
$\tilde{W}_a(\varphi)$ is $a$ independent. 
Then $\tilde{W}_a(\varphi)$ is simply written as 
$\tilde{W}(\varphi)$.
We will prove that $\tilde{W}(\varphi)$ is the majorant of 
$w_a (\varphi)$.
By expanding $\tilde{W}(\varphi)$, $\tilde{F}(\varphi)$
$\tilde{F}(\varphi + \tilde{W}(\varphi))$ as
\begin{equation}
 \tilde{W}(\varphi) = \sum_{|k| \ge 2} W^k \varphi^k,\quad \quad
 \tilde{F}(\varphi) = \sum_{|k| \ge 2} F^k \varphi^k,\quad \quad
 \tilde{F}(\varphi + \tilde{W}(\varphi))  
 = \sum_{|k| \ge 2} D^k \varphi^k,
\end{equation}
we obatin $W^k = D^k / \delta_m$.
For $|k| = 2$, $D^k = F^k$, so $|w^k_a| \le W^k$.
We assume that $|w^k_a| \le W^k$ for $|k| \le K-1$.
Since $d^k_a$ ($|k|=K$) is the 
polynomial of $f^k_a$ ($|k| \le K$) and 
$w^k_a$ ($|k| \le K-1$) with integral coefficients and
$D^k$ is given by substitution $f^k_a \to F^k$, 
$w^k_a \to W^k$ in the polynomial representing $d^k_a$,
$|d^k_a| \le D^k$ ($|k| = K$).
Therefore we obtain $|w^k_a| \le W^k$ ($|k| = K$).
By induction, $\tilde{W}(\varphi)$ is the majorant of $w_a (\varphi)$.
As for the right hand side of (\ref{majorantW}),
\begin{equation}
 \tilde{F}(\varphi + \tilde{W}(\varphi)) = 
 \frac{(\Phi + N_S \tilde{W}(\varphi))^2}
      {1-(\Phi + N_S \tilde{W}(\varphi))},
\quad \quad
 \Phi := \varphi_1 + \cdot \cdot \cdot + \varphi_{N_S},
\end{equation}
so long as $|\Phi + N_S \tilde{W}(\varphi)| < 1$.
So (\ref{majorantW}) can be written as
\begin{equation}
 N_S (N_S + \delta_m) \tilde{W}^2 - 
 [\delta_m - (2 N_S + \delta_m) \Phi] \tilde{W} +
 \Phi^2 = 0,
\end{equation}
whose solution is given by
\begin{equation}
 \tilde{W} = \frac{1}{2 N_S (N_S + \delta_m)}
\left[      
 \delta_m - (2 N_S + \delta_m) \Phi -
 \sqrt{ \delta^2_m - 2 \delta_m (2 N_S + \delta_m) \Phi + \delta^2_m \Phi^2} 
\right]
\end{equation}
for $|\Phi + N_S \tilde{W}(\varphi)| < 1$, using the fact that 
$\tilde{W} = 0$ for $\varphi_a = 0$.
Since $\tilde{W}$ is the majorant of $w_a$, we can prove the convergence
of $w_a$.

This proof in Appendix $J$ is based on \cite{Niwa1981}.

\addtolength{\baselineskip}{-3mm}


\end{document}